\documentclass[%
 reprint,
 amsmath,amssymb,
 aps,
]{revtex4-2}

\usepackage{subfiles}
\usepackage{graphicx}% Include figure files
\usepackage{dcolumn}% Align table columns on decimal point
\usepackage{bm}% bold math
\usepackage{tabularx}
\usepackage{multirow}
\DeclareMathOperator*{\argmax}{arg\,max}
\DeclareMathOperator*{\argmin}{arg\,min}

\newcommand{\ls}{\left[}
\newcommand{\rs}{\right]}
\newcommand{\lp}{\left(}
\newcommand{\rp}{\right)}
\newcommand{\lbr}{\left\{}
\newcommand{\rbr}{\right\}}

\begin{document}

\preprint{APS/123-QED}

% My proposed new title, let me know what you think. I think "Ising solver" is a little played out now, and for an APS journal we may not want to play up "electrical" (I keep flipping back and forth on this.) -PX
% \title{Lagrange Multiplier Binary Optimization with Parametric Oscillators}
% \title{Any coupled parametric oscillator system can perform Lagrange Multiplier primal-dual optimization}
\title{Equivalence of coupled parametric oscillator dynamics to Lagrange multiplier primal-dual optimization}
%\title{A complete Lagrange Multiplier electrical oscillator Ising solver}% Force line breaks with \\

\author{Sri Krishna Vadlamani}
\email{srikv@mit.edu}
\affiliation{Research Laboratory of Electronics, Massachusetts Institute of Technology, Cambridge, Massachusetts 02139, USA
}%

% \collaboration{MUSO Collaboration}%\noaffiliation

\author{Tianyao Patrick Xiao}
\affiliation{Sandia National Laboratories, Albuquerque, New Mexico 87185, USA
}%

\author{Eli Yablonovitch}
\affiliation{Department of Electrical Engineering and Computer Sciences, University of California, Berkeley, California 94720, USA
}%

\date{\today}% It is always \today, today,

\begin{abstract}
There has been a recent surge of interest in physics-based solvers for combinatorial optimization problems. We present a dynamical solver for the Ising problem that is comprised of a network of coupled parametric oscillators and show that it implements Lagrange multiplier constrained optimization. We show that the pump depletion effect, which is intrinsic to parametric oscillators, enforces binary constraints and enables the system's continuous analog variables to converge to the optimal binary solutions to the optimization problem. Moreover, there is an exact correspondence between the equations of motion for the coupled oscillators and the update rules in the primal-dual method of Lagrange multipliers. Though our analysis is performed using electrical LC oscillators, it can be generalized to any system of coupled parametric oscillators. We simulate the dynamics of the coupled oscillator system and demonstrate that the performance of the solver on a set of benchmark problems is comparable to the best-known results obtained by digital algorithms in the literature.
\end{abstract}

\maketitle

%\tableofcontents

\section{\label{sec:level1}Introduction}
% For PRL/PRX, I think it is better to motivate the role of physics from the very first sentence. -PX
There has been significant recent interest in exploiting physical dynamics to solve difficult combinatorial optimization problems. While these problems are of great interest to several application domains, many are \textbf{NP}-hard \cite{karp_reducibility_1972}. Conventional digital algorithms are either based on provable approximations to provide a lower bound on solution quality \cite{goemans1995improved}, or on heuristics (and metaheuristics) to search for higher-quality solutions \cite{benlic_breakout_2013}. Alternatively, these problems can potentially be solved faster and more efficiently by embedding an optimization algorithm in the dynamical equations of a physical system. Such a system must be both physically realizable and capable of finding comparable or superior quality solutions to state-of-the-art digital algorithms.

Physics-based optimization machines are often designed to solve the Ising problem, relying on the fact that any $\textbf{NP}$-hard problem can be reduced to the Ising problem in polynomial time \cite{lucas_ising_2013}. The Ising problem consists of a set of $N$ interacting spins, each of which has two possible orientations, and the challenge is to find the configuration of spins $\bm{x}$ that minimizes the total interaction energy $f$, given by:
\begin{equation}
    f(\bm{x})=-\sum_{i=1}^N\sum_{j=1}^N J_{ij}x_ix_j
    \label{eq:Ising}
\end{equation}
where $x_i = \pm 1$ is the binary orientation of the $i^\text{th}$ spin, and $J_{ij}$ is the interaction strength between the $i^\text{th}$ and $j^\text{th}$ spins. Minimizing the Ising energy has a one-to-one equivalence to the MAXCUT problem, whose instances are typically used to benchmark the algorithmic performance of Ising solvers.

%$\bm{x}$ is the vector whose components are the $x_i$ for all the spins, and the interaction energy between the $i$-th and $j$-th spins is given by $-J_{ij}x_ix_j$ for a given interaction strength coefficient $J_{ij}$. The Ising problem asks us to find the vector $\bm{x}$ that minimizes the Ising energy \eqref{Ising}.

% Instead of listing them all out then classifying them, it would be much clearer to introduce the different references by class
% 3/17: This paragraph still needs extensive modifications/re-write
Physical Ising solvers take many forms in the literature, including quantum \cite{Dwave2011} and classical machines \cite{goto_combinatorial_2019,goto_quantum_2019} that utilize the adiabatic principle. Many other classical hardware solvers adopt a Hopfield neural network or Boltzmann machine approach to minimize the Ising energy\textemdash they perform iterative matrix operations to update the binary variables and implement simulated annealing \cite{van1987simulated,Cai2020,Bojnordi2016,Strukov2019,Suhas2017, patel2022logically,roques-carmes_heuristic_2020,pierangeli_large-scale_2019}. There are approaches such as Memcomputing \cite{traversa_polynomial-time_2017,di_ventra_perspective_2018}, chaotic dynamical systems \cite{leleu_destabilization_2019, molnar_continuous-time_2018, ercsey-ravasz_optimization_2011}, and a number of approaches based on coupled bistable dynamical elements, such as stochastic magnetic bits \cite{camsari2017stochastic,borders_integer_2019} or coupled oscillators. Many types of coupled oscillators have been proposed: laser parametric oscillators \cite{wang_coherent_2013,marandi_network_2014,inagaki_coherent_2016}, injection-locked LC oscillators \cite{mcquillan_oim_2019,Chou2019}, CMOS ring oscillators \cite{Ahmed2021}, phase-transition oscillators \cite{Dutta2021}, coupled multicore fiber lasers \cite{babaeian_single_2019}, and coupled polaritonic cavities \cite{kalinin_global_2018}. Each oscillator is forced into bistability by a physical nonlinearity, and chooses one of the two stable states based on the strengths of its interactions with other oscillators.

While combinatorial optimization problems can be mapped to and solved by coupled oscillator networks, the algorithm implemented by these physical systems is not always clearly understood from the viewpoint of optimization theory. A key consideration is how the constraint of binary variables is imposed on the physical system. It was previously shown that the coupled oscillator systems were implementing the primal part of the gradient-based primal-dual method of Lagrange multipliers \cite{vadlamani2020physics}. In that work, it was proposed that auxiliary hardware apart from the oscillators was needed to implement the dual portion of the algorithm\textemdash appropriately controlling the system's analogue of the Lagrange multipliers in order to strictly impose binary fixed-amplitude constraints ($\pm 1$ for the Ising spins) on the problem. The need to impose binary fixed-amplitude constraints is recognized as the `problem of imposing amplitude homogeneity' by Leleu et al. \cite{leleu_destabilization_2019}. 

In this paper, we show that the dual portion of the Lagrange multiplier method\textemdash the evolution of the Lagrange multipliers \textemdash is automatically implemented by the phenomenon of pump depletion within each parametric oscillator. Pump depletion provides the necessary feedback to constrain the oscillator states into amplitude-stability and phase-bistability, without need for any auxiliary hardware. The equations of motion of the signal and pump oscillators implement exactly the alternating primal and dual steps of the Lagrange multiplier method. We exploit this phenomenon to fully map the method of Lagrange multipliers onto the dynamics of parametric oscillators, and show that these dynamics can find high-quality solutions to the MAXCUT problem. Furthermore, we find that the solution quality is robust to imprecision in the coupling components, thus circumventing a historical shortcoming of analog solvers.

% Split on the following sentence which I wrote. Include this if we want to emphasize electrical for this paper. We definitely want to do this for a venue like IEEE but maybe not APS.
%The dynamics are derived from the circuit equations for coupled LC parametric oscillators, but take a form that can be generalized to other physical implementations of parametric oscillators.

% These sentences absolutely kill the flow of the intro. Please move them later in the paper
%We comment at this point that the non-gradient based methods of Ercsey-Ravasz et al. \cite{ercsey-ravasz_optimization_2011}, \cite{molnar_continuous-time_2018}, and Leleu et al. \cite{leleu_destabilization_2019} do not fall within the category of Lagrange multiplier optimization. These algorithms have built-in elements of Lagrange multipliers but also incorporate non-gradient dynamics that modifies their behavior. 

The paper is organized as follows: Section 2 presents the equations of motion for a network of electrical parametric LC oscillators, which acts as a proxy for all the coupled oscillator approaches, and shows that pump depletion can be used to constrain the oscillators to fixed-amplitude binary states. Section 3 briefly reviews the primal-dual Lagrange multiplier method for constrained optimization. Section 4 shows that for the $\pm 1$ binary constraint, there is a complete and exact equivalence between the oscillator dynamics and the Lagrange multiplier method, including the Augmented Lagrange method. In Section 5, we numerically simulate the equations of motion of the electrical oscillators, which implement the Augmented Lagrange method, and compare the results with two other algorithms on several large MAXCUT problems in the Gset and BiqMac problem sets \cite{gset, wiegele2007biq}. Section 6 concludes the paper.

\section{Ising Energy Minimization With Coupled Parametric LC Oscillators}
\label{sec:parametric_oscillators}

We map an interacting ensemble of $N$ Ising spins to a network of $N$ resistively coupled parametric LC oscillators. This system was first discussed by Xiao \cite{xiao2019optoelectronics} and its connections to Lagrange multipliers were explored by Vadlamani et al. \cite{vadlamani2020physics} and Vadlamani \cite{vadlamani2021sharp}. Parametric amplification, enabled by a capacitive nonlinearity in the LC oscillator, ensures that each oscillator's steady state is bistable in phase indicating that these systems can be used to implement binary Ising spins. However, to fully implement $\pm 1$ spins, one also needs amplitude-stability in addition to phase-bistability. We show how both these conditions are achieved in this section.

%The coupled LC oscillator solver is composed of several parametric LC oscillators that are resistively coupled to one another. The capacitor voltage in each oscillator in the network represents a spin in the specified Ising problem\textemdash an $N$-spin problem is implemented by $N$ oscillators. To enable the capacitor voltages to represent the binary states of the spin, we use parametric amplification to restrict each oscillator to one global quadrature of oscillation. The dynamics of the parametric pump serves to further restrict the oscillators to a fixed amplitude within the chosen quadrature. To make the mapping between the physical system and the Ising problem exact mathematically, the capacitor voltages should settle down to predetermined fixed amplitudes along the chosen quadrature; simple collapse of the oscillators to a single quadrature is not enough.   

\subsection{Mapping Ising spins to parametric oscillators}
\begin{figure}[t]
\centering
\includegraphics[width=0.5\textwidth]{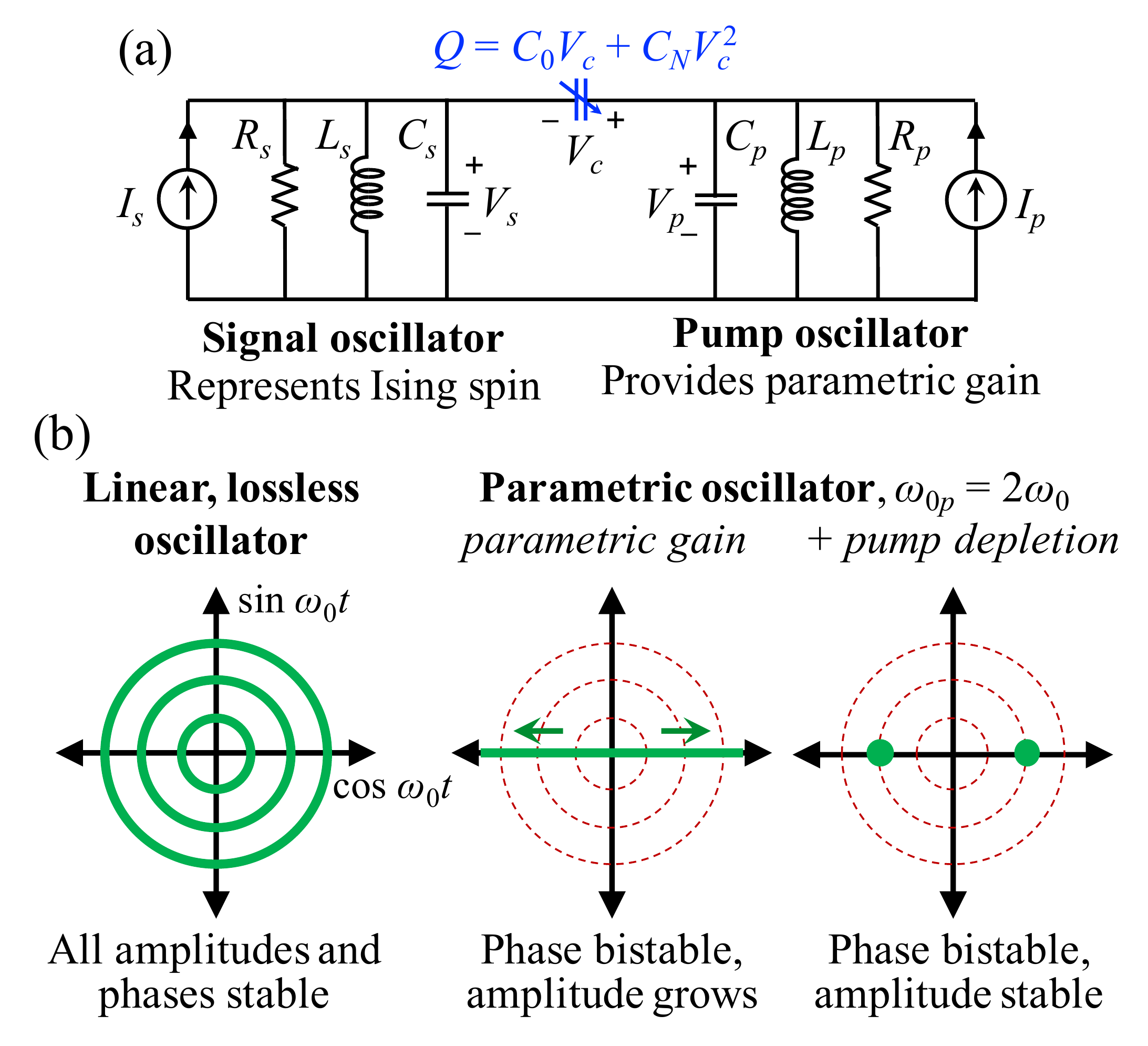}
\caption{(a) Parametric LC oscillator circuit, consisting of a signal and pump oscillator connected by a nonlinear capacitor. (b) A linear, lossless LC oscillator (left) can support oscillations of any phase and any amplitude. A parametric LC oscillator (right) supports oscillations only at two phases separated by $\pi$ rad as a result of parametric amplification. Pump depletion further constrains the amplitude to be monostable.}
\label{fig:parametric}
\end{figure}

A linear LC oscillator supports sinusoidal oscillations $A\cos{\lp\omega_0t+\phi\rp}$ of arbitrary amplitude $A$ and phase $\phi$, where $\omega_0=1/\sqrt{LC}$ is the natural frequency of the LC cavity. To constrain the oscillator's phase and amplitude, we induce parametric amplification; a parameter of the oscillator is modulated by a second oscillator at $2\omega_0$, called the pump. We choose the capacitance as the modulated parameter, and enable its modulation by introducing a second-order nonlinear capacitance. The parametric oscillator circuit, from \cite{parametricjpnotes}, is shown in Fig. \ref{fig:parametric}(a), where the nonlinear capacitor couples the original oscillator (called the signal) and the pump oscillator. The nonlinear capacitor's characteristic is:
\begin{equation}
    Q=C_0V_c+C_NV_c^2\label{capnonlin1}
\end{equation}
where $C_0$ is the linear capacitance and $C_N$ is the second-order nonlinear capacitance. The second term can be viewed as a capacitance that is modulated by the voltage $V_c$, which depends on both the pump voltage $V_p$ and the signal voltage $V_s$. The nonlinear capacitance can be implemented by common semiconductor devices such as $p$-$n$ junction or Schottky diodes.

%\begin{figure}[h]
%\centering
%\includegraphics[scale = 0.37]{PhaseBistability.pdf}
%\caption{Parametric pumping induces bistability in phase in LC oscillators. A normal LC %oscillator, shown on the left, can support oscillations of any phase. A parametric LC oscillator, shown on the right, can support oscillations only at two phases, $\phi_0$ and $\phi_0+\pi$, for some $\phi_0$.}
%\label{fig:PhaseBistab}
%\end{figure}

To induce phase bistability, the pump acts to modulate the nonlinear capacitance at twice the resonance frequency of the signal oscillator. If the signal's power peaks while the capacitance falls, energy is transferred from the pump to the signal oscillator, providing gain: this occurs for two specific phases of the signal oscillator, separated by $\pi$ radians. If this parametric gain exceeds the signal oscillator's resistive losses, the amplitude grows. Meanwhile, for the other phase quadrature, energy flows from the signal to the pump, and the amplitude decays. As a result, only one quadrature survives and the oscillator becomes phase bistable, as shown in Fig. \ref{fig:parametric}(b) (middle).

Phase bistability is not sufficient to implement binary spins; the oscillation amplitudes must also be stable. When parametric gain is first introduced, the oscillator amplitude increases exponentially with time. As the signal amplitude increases, it draws more power from the pump to sustain its growth. This continues until the pump depletes the finite amount of power supplied to it (modeled in Fig. \ref{fig:parametric}(a) as a constant current source). The signal and pump then exchange power back and forth until both amplitudes settle around a steady-state value. This mechanism allows the oscillator to be truly bistable, as shown in Fig. \ref{fig:parametric}(b) (right).

To derive the equations of motion for the parametric oscillator, we solve Kirchoff's circuit equations for the circuit in Fig. \ref{fig:parametric}(a). This is shown in Appendix \ref{appsec1}. A key step in the derivation is the use of the slowly-varying amplitude approximation, which assumes that the amplitude envelopes ($A_s$ and $A_p$) of the oscillating voltages ($V_s$ and $V_p$) vary slowly compared to the frequency of the harmonic oscillations themselves. Under this approximation, the amplitude envelopes can be shown to evolve as:
%In Appendix 1, we expand the signal voltage as $V_s(t)~=~A_s(t)\cos\lp\omega_0t+\phi\rp~+~B_s(t)\sin\lp\omega_0t+\phi\rp$, and the pump voltage as $V_p(t)=A_p(t)\cos\lp2\omega_0t\rp$. The approximation involves assuming that the rate of change of the envelope functions $A_s(t),B_s(t),A_p(t)$ is small compared to the oscillation frequency $\omega_0$. The detailed calculation, provided in the appendix, yields the following equations of motion for the voltage envelope functions:
\begin{align}
\label{eq:As}
%    \frac{dA_s}{dt}&=\frac{I_s}{2\lp C_0+C_s\rp}-\frac{A_s}{2R_s\lp C_0+C_s\rp}+\frac{C_N\omega_0A_sA_p}{2\lp C_0+C_s\rp}\\
        \frac{dA_s}{dt}&=\frac{1}{2}\left[\frac{I_s}{C_0+C_s}-\frac{A_s}{R_s\lp C_0+C_s\rp}+\frac{C_N\omega_0A_sA_p}{C_0+C_s}\right]\\
\label{eq:Ap}
    \frac{dA_p}{dt}&=\frac{1}{2}\left[\frac{I_p}{C_0+C_p}-\frac{A_p}{R_p\lp C_0+C_p\rp}-\frac{C_N\omega_0A_s^2}{C_0+C_p}\right]
\end{align}
%where $A_s$ and $A_p$ are the slowly-varying amplitudes of voltage across the capacitors in the signal and pump circuits respectively, the other symbols have the same meanings they have in Fig. \ref{}, and $\phi$ was set to $3\pi/4$ to get convenient equations. The sine component $B_s$ is dropped altogether because it experiences loss due to the interaction with the pump and doesn't achieve an appreciable amplitude. A more detailed justification for ignoring $B_s$ is provided in the appendix. 
%Equations \eqref{eq:As} and \eqref{eq:Ap} are both very intuitive. 

While the equations above were derived for the specific nonlinear LC oscillator circuit, they have a general form that can describe many different types of parametric oscillators consisting of a pump and signal oscillator. In both equations, the first term is a power source, the second term corresponds to internal dissipation, and the third term represents the exchange of power between the pump and signal. The signal $A_s$ has a parametric gain term that is proportional to the pump amplitude, while the pump $A_p$ has a loss term corresponding to the transfer of energy to the signal oscillator. This term is responsible for pump depletion, which limits the parametric gain and the signal amplitude. In the following, we assume that the signal oscillator does not have its own power source and instead, $I_s$ corresponds to noise power with a time-averaged current of zero.

%The signal equation has an input term, $I_s$, a resistive decay term due to the internal resistance, and a gain term that is directly proportional to $V_p$, the pump capacitor voltage. The pump equation has an input term, $I_p$, a resistive decay term due to the internal resistance, and a `pump depletion' term that describes the loss of the pump amplitude in response to an excessive increase in the signal amplitude. 

%In the next subsection, we construct an Ising solver by coupling several of these parametric oscillators via resistive connections.

% It would be good to add a paragraph here saying that the equations for A_s and A_p are identical in form to equations for laser systems, with references. Sri I think you know these references better than me. -PX

\subsection{Dynamics of dissipatively coupled parametric oscillators}
\label{subsec:coupled_dynamics}

\begin{figure}[t!]
\centering
\includegraphics[scale = 0.47]{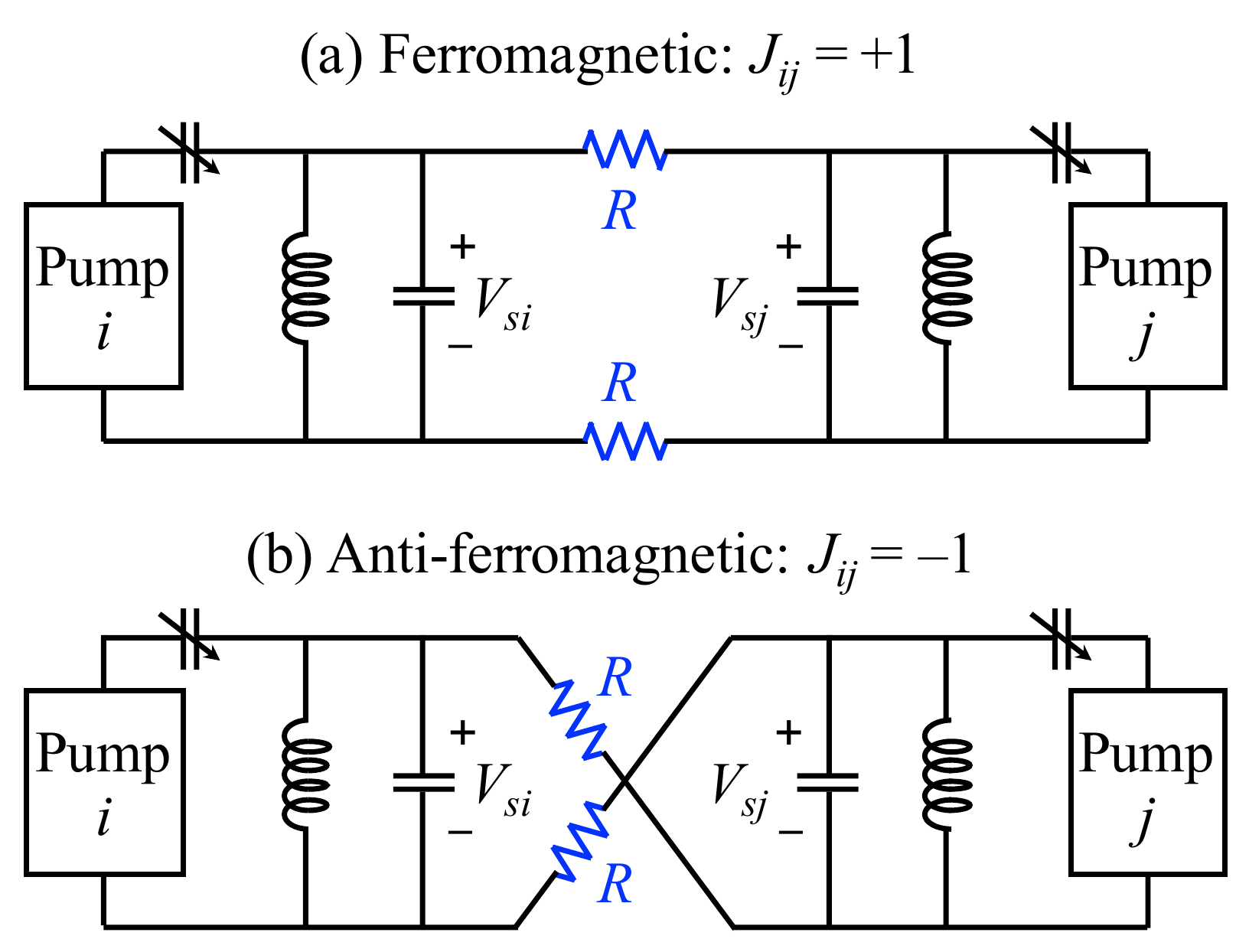}
\caption{Scheme for (a) ferromagnetic and (b) anti-ferromagnetic coupling of two LC oscillators.}
\label{fig:coupling}
\end{figure}

Fig. \ref{fig:coupling} shows a scheme to resistively couple the bistable LC oscillators to implement the spin-spin interactions $J_{ij}$ in the Ising problem, also used previously by Wang \textit{et al.} \cite{mcquillan_oim_2019}. For simplicity, we consider the case where the interaction weights are binary: $J_{ij} = \pm 1$, but the scheme can straightforwardly be extended to any intermediate-valued weights (see Appendix \ref{appsec2b}).

% Would it be worth adding a sentence at the end of this paragraph about the Onsager relations? -PX
If two oscillators are ferromagnetically coupled ($J_{ij}=1$), a pair of straight-linking resistors induces them to oscillate together with the same phase. If they start with opposite phases, the large voltage differences across the connection resistors cause a flow of current that can flip the phase of an oscillator. Conversely, for two anti-ferromagnetically coupled oscillators ($J_{ij}=-1$), a pair of cross-linking resistors are used. A frustrated spin interaction dissipates more power; by dissipatively coupling the oscillators, the network evolves toward a state that minimizes the collective power dissipation.

% I went ahead and plugged in \sum |J_ij| in place of N-1 everywhere since there's no reason to assume all-to-all coupling
% I also got rid of the second equation because it's redundant with Eq 4. Just added a sentence on it.
The equations of motion for the full network of $N$ coupled, identical parametric LC oscillators are derived from Kirchoff's circuit laws. This is shown in Appendix \ref{appsec2}, assuming the slow-varying amplitude approximation. The result modifies the equations of motion for the $i^\text{th}$ oscillator to account for spin-spin interactions:
\begin{align}
\label{eq:Asi}
    \frac{dA_{si}}{dt}&=-\ls\frac{N_i A_{si}}{4RC_{0s}} -\frac{1}{4RC_{0s}}\sum_{j:j\ne i}J_{ij} A_{sj}\rs+\frac{C_N\omega_0A_{pi}A_{si}}{2C_{0s}}
%\label{eq:Api}
%    \frac{dA_{pi}}{dt}&=\frac{I_{p}}{2\lp C_0+C_p\rp}-\frac{C_N\omega_0 A_{si}^2}{2\lp C_0+C_p\rp}
\end{align}
where $C_{0s}=C_0+C_s$, $R$ is the coupling resistance, and $N_i := \sum_{j:j\ne i} |J_{ij}|$ is the number of nonzero connections to the $i^\text{th}$ oscillator. 

The form of this equation can be generalized to any coupled parametric oscillator network. The term in square brackets in Eq. \eqref{eq:Asi} captures the net loss that the signal amplitude $A_{si}$ experiences due to its connections to the other oscillators. The last term is the parametric gain that is supplied from the pump, as in Eq. \eqref{eq:As}. We have temporarily assumed that  the internal dissipation within each oscillator is negligible compared to the loss in the oscillator connections.

%The extension to general coupling is straightforward and leads to $\sum_j|J_{ij}|$ instead of $(N-1)$ in all the expressions (see Appendix \ref{}). As before, Eqs. \eqref{} and \eqref{} have straightforward physical interpretations. 

We now re-express the system's dynamics to better elucidate its algorithmic functionality. First, the pump dynamics in Eq. \eqref{eq:Ap} can be re-written as:
\begin{align}
    \label{eq:dApi_dt}
    %\frac{dA_{pi}{d}&=\frac{C_N\omega_0}{2\lp C_0+C_p\rp}\lp\frac{I_{p}}{C_N\omega_0}-A_{si}^2\rp\\
    \frac{dA_{pi}}{dt} = \frac{C_N\omega_0}{2\lp C_0+C_p\rp}\lp A_{\text{sat}}^2-A_{si}^2\rp
\end{align}
where we have introduced $A_{\text{sat}}^2~:=~I_{p}/C_N\omega_0$. We have assumed that $R_p$ is large enough that the pump's internal loss quickly becomes negligible relative to the loss due to the transfer of energy to the signal oscillator. 

Next, by incorporating the pump amplitude into a new variable,
\begin{equation}
\label{eq:circuit_lambda}
    \Lambda_i:=\frac{C_N\omega_0A_{pi}}{2}-\frac{N_i}{4R}
\end{equation}
we observe that the dynamics for $A_{si}$ and $A_{pi}$ can be re-expressed as follows:
\begin{align}
% I have decided to take out the intermediate equations of this derivation which I don't think add much value
\label{eq:Asi_lagrange}
%    \frac{dA_{si}}{dt}&=\frac{1}{4RC}\sum_{j:j\ne i}J_{ij} A_{sj}+\frac{\lambda_i A_{si}}{C} = -\frac{1}{2C} \frac{\partial L}{\partial A_{si}} \\[2ex]
    \frac{dA_{si}}{dt}&= -\frac{1}{2C_{0s}} \, \frac{\partial \mathcal{L}}{\partial A_{si}} \\[2ex] %&=-\frac{1}{2C}\frac{\partial}{\partial A_{si}}\ls-\frac{1}{4R}\sum_{j,k}J_{jk} A_{sj}A_{sk}+\sum_k\lambda_k\lp A_{\text{sat}}^2-A_{sk}^2\rp\rs\label{netsignal}\\
\label{eq:lambdai_lagrange}
%    \frac{d\lambda_i}{dt}&=\frac{C_N^2\omega_0^2 \lp A_{\text{sat}}^2-A_{si}^2\rp}{4\lp C_0+C_p\rp} = \frac{C_N^2\omega_0^2}{4\lp C_0+C_p\rp} \frac{\partial L}{\partial \lambda_i}
    \frac{d\Lambda_i}{dt}&=\frac{C_N^2\omega_0^2}{4\lp C_0+C_p\rp}\lp A^2_{\text{sat}}-A^2_{si}\rp= \frac{C_N^2\omega_0^2}{4\lp C_0+C_p\rp} \,\frac{\partial \mathcal{L}}{\partial \Lambda_i}
%    &=\frac{C_N^2\omega_0^2}{4\lp C_0+C_p\rp}\frac{\partial}{\partial \lambda_i}\ls-\frac{1}{4R}\sum_{j,k}J_{jk} A_{sj}A_{sk}+\sum_k\lambda_k\lp A_{\text{sat}}^2-A_{sk}^2\rp\rs\label{netpump}
\end{align}
where $\mathcal{L}$ is defined as:
\begin{equation}
\label{eq:ckt_lagrange}
    \mathcal{L}\lp \bm{A_s},\bm{\Lambda}\rp=-\frac{1}{4R}\sum_{i,j}J_{ij} A_{si}A_{sj}+\sum_i\Lambda_i\lp A_{\text{sat}}^2-A_{si}^2\rp
\end{equation}
with $\bm{A_s}$ and $\bm{\Lambda}$ being vectors whose components are $A_{si}$ and $\Lambda_i$ respectively. We call the quantity $\mathcal{L}$ the \textbf{Lagrange function} of the problem. The above equations show that the signal and pump amplitudes respectively perform simultaneous gradient descent and ascent on the same function $\mathcal{L}$.

Notably, the first term in Eq. \eqref{eq:ckt_lagrange} has the form of the Ising interaction energy in Eq. \eqref{eq:Ising}, except that the amplitudes are not strictly binary. According to Equation \eqref{eq:Asi_lagrange}, the amplitudes of the oscillators evolve to minimize this Ising-like function. However, this does not fully describe the dynamics, due to the presence of the second term in the Lagrange function. We will show that these equations of motion are actually an exact implementation of the primal-dual method of Lagrange multipliers. In the next section, we provide a brief overview of the method of Lagrange multipliers, and in Sec. \ref{sec:correspondence}, we make the isomorphism between the circuit and the Lagrange multiplier method more explicit.

\section{Lagrange Multipliers Overview}
The method of Lagrange multipliers is a well-known procedure for solving constrained optimization problems. Here, we provide a brief overview of the method, and refer the reader to Bertsekas \cite{bertsekas_nonlinear_1999} and Boyd and Vandenberghe \cite{boyd_convex_2004} for further details.

\begin{figure}[t]
\centering
\includegraphics[scale = 0.43]{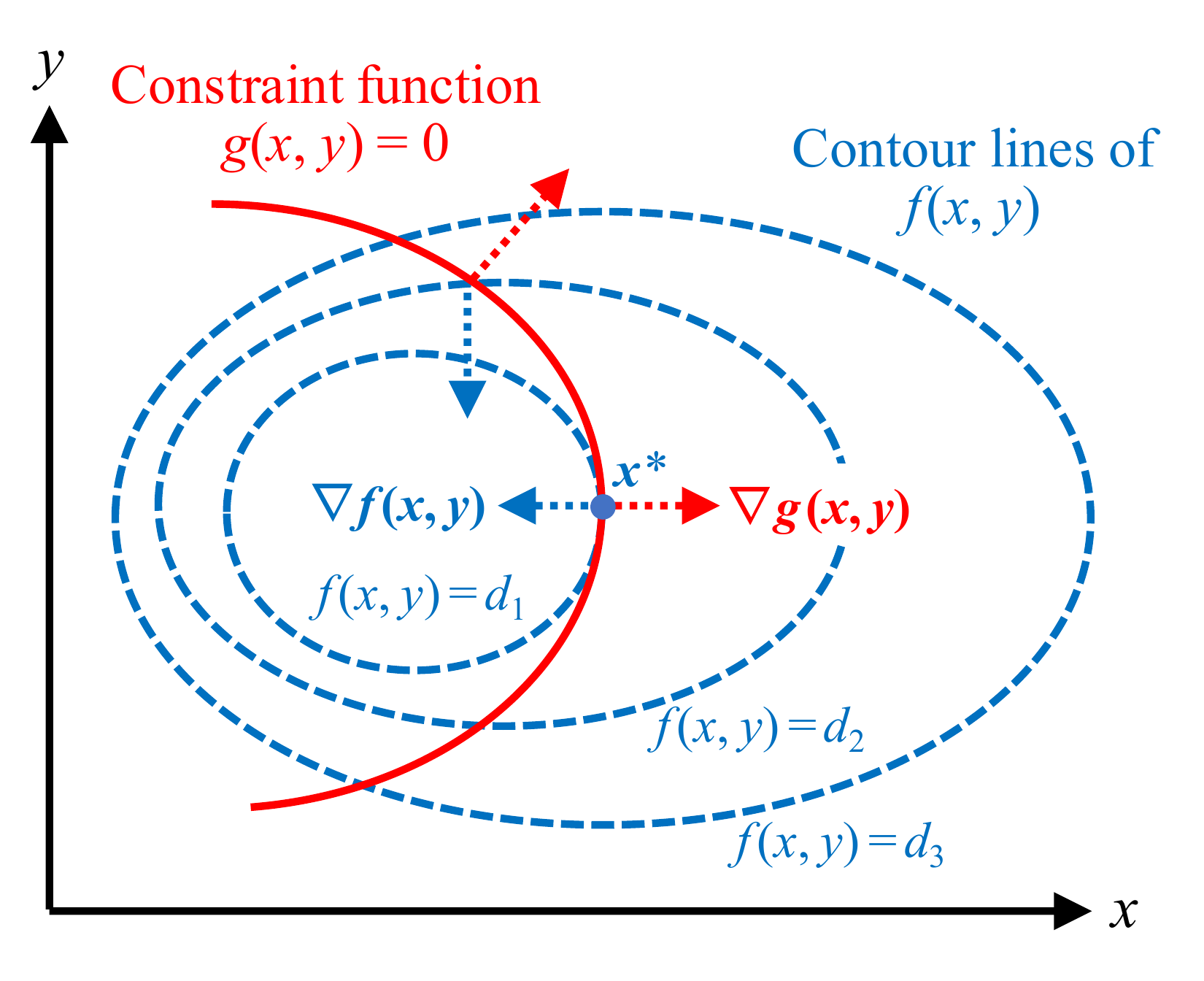}
\caption{Maximization of function $f(x,y)$ subject to the constraint $g(x,y)=0$. At the constrained local optimum, the gradients of $f$ and $g$ are parallel.}
\label{fig:lag}
\end{figure}

Let $f(\bm{x})$ be a merit function of $N$ variables, and let $\bm{x^*}$ be a point that locally minimizes $f(\bm{x})$ amongst the set of all $\bm{x}$ that satisfy a given constraint $g(\bm{x})=0$. 
%That is, $\bm{x^*}$ is a constrained local minimum of $f(\bm{x})$ and the merit function remains unchanged for small deviations around $\bm{x^*}$ that respect the constraints.
That is, $f(\bm{x})$ does not change when one makes infinitesimal displacements about $\bm{x^*}$ that are tangential to the constraint curve $g(\bm{x})=0$. This means $\bm{\nabla}f(\bm{x^*})$ and $\bm{\nabla}g(\bm{x^*})$ should be parallel to each other:
\begin{equation}
    \bm{\nabla} f(\bm{x^*})=-\lambda^*\bm{\nabla} g(\bm{x^*})\label{singconst}.
\end{equation}
The proportionality constant $\lambda^*$ is called the Lagrange multiplier corresponding to the constraint $g(\bm{x})=0$. A two-dimensional example for maximization is shown in Fig.~\ref{fig:lag}. When there are multiple constraints $g_1=0,\dots,g_p=0$, Eq. \eqref{singconst} is generalized as follows:
\begin{equation}
    \bm{\nabla} f(\bm{x^*})=-\sum_{i=1}^p\lambda_i^*\bm{\nabla} g_i(\bm{x^*})\label{multiconst},
\end{equation}
Every point $\bm{x^*}$ that renders $f(\bm{x})$ locally stationary subject to the constraints satisfies Eq. \eqref{multiconst} for some $\bm{\lambda^*}$, where $\bm{\lambda^*}$ is the vector whose components are $\lambda_i^*$. We now define a Lagrange function:
\begin{equation}
    L(\bm{x}, \bm{\lambda}) := f(\bm{x}) + \sum_{i=1}^p\lambda_ig_i(\bm{x}),\label{lagrangefunc}
\end{equation}
This function has the property that any stationary point $\bm{x^*}$ and its associated multipliers $\bm{\lambda^*}$ satisfy:
\begin{equation}
    \bm{\nabla_x}L(\bm{x^*},\bm{\lambda^*})=0,\ \bm{\nabla_\lambda}L(\bm{x^*},\bm{\lambda^*})=0.
    \label{eq:KKT}
\end{equation}
% Is it necessary to name this for our purposes? -PX
%These are the Karush-Kuhn-Tucker (KKT) necessary and sufficient conditions for local stationarity. 
If a candidate point $\lp\bm{x'},\bm{\lambda'}\rp$ satisfies these conditions, then $\bm{x'}$ is a stationary point of $f(\bm{x})$ subject to the constraints. Eq. \eqref{eq:KKT} transforms the problem of finding constrained stationary points of $f(\bm{x})$ to that of finding unconstrained stationary points of $L(\bm{x},\bm{\lambda})$.
%The second-order KKT conditions, which involve the second derivatives of $L(\bm{x},\bm{\lambda})$ at $\lp\bm{x'},\bm{\lambda'}\rp$, can be used to decide if the stationary point $\bm{x'}$ is specifically a local minimum , a local maximum, or a saddle point of $f(\bm{x})$ subject to the constraints.   

%\subsection{Algorithm to find constrained optimal $\bm{x^*}$ and its associated multipliers $\bm{\lambda^*}$}

% In general, a stationary point $\bm{x^*}$ and its associated multipliers $\bm{\lambda^*}$ form a saddle point of $L(\bm{x},\bm{\lambda})$. 

%With the aim of motivating a well-known iterative algorithm, called the `Method of Multipliers' \cite{bertsekas_nonlinear_1999}, to find $\lp\bm{x^*},\bm{\lambda^*}\rp$, we now narrow our focus temporarily to special optimization problems that satisfy a property called `strong duality'. The Ising problem does not satisfy this property so we shall modify the iterative algorithm to make it applicable to general optimization problems at the end of this subsection. 

Certain `well-structured' problems (e.g. convex problems) satisfy `strong duality':
\begin{equation}
    \min_{\bm{x}}\lp\max_{\bm{\lambda}}L(\bm{x},\bm{\lambda})\rp= \max_{\bm{\lambda}}\lp\min_{\bm{x}}L(\bm{x},\bm{\lambda})\rp\label{strongdualmain}
\end{equation}
The point where the equality holds is the global constrained optimum $\lp\bm{x^*},\bm{\lambda^*}\rp$ of the problem. The Method of Multipliers \cite{bertsekas_nonlinear_1999} finds this optimum by solving the nested min-max optimization problem on the right-hand side of Eq. \eqref{strongdualmain} iteratively. Starting from a point $(\bm{x^{(0)}},\bm{\lambda^{(0)}})$, the algorithm first keeps $\bm{\lambda}$ fixed and minimizes $L$ using several gradient descent steps in $\bm{x}$. Then, $\bm{x}$ is kept fixed and one step of gradient ascent on $L$ is performed in the $\bm{\lambda}$ directions. This alternating fast minimization-slow maximizaton procedure is repeated until convergence. In the limit of zero step size, the iterative algorithm can be converted into a pair of differential equations in time:
\begin{align}
\label{eq:MM1}
    \frac{d\bm{x}}{dt}&=-\kappa\bm{\nabla_x}L(\bm{x},\bm{\lambda})\\
\label{eq:MM2}
    \frac{d\bm{\lambda}}{dt}&=\kappa'\bm{\nabla_{\lambda}}L(\bm{x},\bm{\lambda})
\end{align}
for suitably chosen stepsizes $\kappa$ and $\kappa'$. Eq. \eqref{eq:MM1} corresponds to gradient descent on the Lagrange function to optimize $\bm{x}$, while Eq. \eqref{eq:MM2} corresponds to gradient ascent  on $L$ to optimize $\bm{\lambda}$. This procedure is also called the primal-dual algorithm \cite{goemans1997primal}, where the descent in $\bm{x}$ is the primal step and the ascent in $\bm{\lambda}$ is the dual step. Strong duality guarantees that the algorithm converges on the global optimum of $f$ that satisfies the constraints. By the nature of Eq. \eqref{strongdualmain}, the algorithm can also proceed by performing a fast maximization over $\bm{\lambda}$ in conjunction with a slow minimization over $\bm{x}$, thereby solving the left-hand side formulation.

Unfortunately, most difficult problems are highly non-convex and only satisfy weak duality: 
\begin{equation}
    \min_{\bm{x}}\lp\max_{\bm{\lambda}}L(\bm{x},\bm{\lambda})\rp\geq \max_{\bm{\lambda}}\lp\min_{\bm{x}}L(\bm{x},\bm{\lambda})\rp\label{weakdual}
\end{equation}
In this case, performing a fast maximization over $\bm{\lambda}$ in conjunction with a slow minimization over $\bm{x}$ is more accurate since the left hand side of Eq. \eqref{weakdual} is in fact the required constrained minimum. 
Alternatively, one could use the more powerful Augmented Lagrangian Method of Multipliers \cite{bertsekas_nonlinear_1999} when strong duality is not satisfied. The method defines an Augmented Lagrange function, $L_{\alpha}(\bm{x},\bm{\lambda})$:
\begin{align}
    L_{\alpha}(\bm{x},\bm{\lambda}) := L(\bm{x},\bm{\lambda})+\frac{\alpha}{2}\left(\sum_{i=1}^p\left(g_i(\bm{x})\right)^2\right)
\end{align}
for a positive parameter $\alpha$. Bertsekas \cite{bertsekas_nonlinear_1999} shows that if the $L(\bm{x},\bm{\lambda})$ in Eqs. \eqref{eq:MM1} and \eqref{eq:MM2} is replaced with $L_{\alpha}(\bm{x},\bm{\lambda})$ and the system is initialized close to a local optimum $\lp\bm{x^*},\bm{\lambda^*}\rp$, the equations will converge to $\lp\bm{x^*},\bm{\lambda^*}\rp$.

%The new, more general differential equations are:
%% I think these equations can probably be cut since they're practically identical to the ones above -PX
%\begin{align}
%    \frac{d}{dt}\bm{x}&=-\kappa\bm{\nabla_x}L_c(\bm{x},\bm{\lambda})\label{auglagx}\\
%    \frac{d}{dt}\bm{\lambda}&=\kappa'\bm{\nabla_{\lambda}}L_c(\bm{x},\bm{\lambda})\label{auglaglam}
%\end{align}
%In the next section, we show that the equations of the motion of the coupled oscillator network, Eqs. \eqref{} and \eqref{}, exactly match the Eqs. \eqref{} and \eqref{} for the Ising problem. Moreover, we also show that the incorporation of nonlinear resistors in the circuits enables an exact implementation of the Augmented Lagrange equations of motion Eqs. \eqref{auglagx} and \eqref{auglaglam}.  

\section{Signal dynamics performs primal step, pump dynamics performs dual step}
\label{sec:correspondence}

We now apply the method of Lagrange multipliers to the Ising optimization problem, whose merit function $f(\bm{x})$ is given in Equation \eqref{eq:Ising}, with the constraint that each of the $N$ spins is binary: $x_i = +1$ or $x_i = -1$. This binary constraint can be written as: $g_i(\bm{x}) = 1 - x_i^2 = 0$ for all $i$ from 1 to $N$. The Lagrange function for the Ising problem is then given by:
\begin{equation}
\label{eq:IsingLagrange}
    L(\bm{x},\bm{\lambda})=-\sum_{i=1}^N\sum_{j=1}^NJ_{ij}x_ix_j+\sum_{i=1}^N\lambda_i\lp1-x_i^2\rp
\end{equation}
where $\lambda_i$ is the Lagrange multiplier associated with the constraint on the $i^\text{th}$ spin. Substituting this expression into Eqs. \eqref{eq:MM1} and \eqref{eq:MM2}, we derive the update equations for the primal-dual Method of Multipliers:
\begin{align}
\label{eq:MMx}
    \frac{dx_i}{dt} &=-2\kappa\left(-\sum_{j=1}^NJ_{ij}x_j-\lambda_ix_i\right)\\
\label{eq:MMlambda}
    \frac{d\lambda_i}{dt} &=\kappa'\lp1-x_i^2\rp
\end{align}

Notably, the Ising problem's Lagrange function in Eq. \eqref{eq:IsingLagrange} is in exact correspondence to the quantity $\mathcal{L}$ we had defined as the Lagrange function of the coupled oscillator network in Section \ref{sec:parametric_oscillators}. Moreover, the equations of motion of the Method of Multipliers, Eqs. \eqref{eq:MMx} and \eqref{eq:MMlambda}, are in perfect correspondence with the oscillator network's equations of motion, Eqs. \eqref{eq:Asi_lagrange} and \eqref{eq:lambdai_lagrange}. More precisely, the signal equation \eqref{eq:MMx} exactly corresponds to the primal equation \eqref{eq:Asi_lagrange} while the pump equation \eqref{eq:MMlambda} exactly corresponds to the dual equation \eqref{eq:lambdai_lagrange}. One only needs to make the identifications given in Table \ref{tab:mapping} to complete the correspondence. Besides $A_{si}$, $\Lambda_i$, and $\mathcal{L}$, all other physical parameters in Table \ref{tab:mapping} are fixed constants. Therefore, the coupled oscillator circuit in fact implements the two differential equations that describe the primal-dual Method of Multipliers.

We make a comment about pump loss here. Since Eq. \eqref{eq:lambdai_lagrange} was obtained by assuming that the pump was lossless, the correspondence is exact when the circuit is close to this regime. A discrepancy arises between the two solvers when the pump loss is too large\textemdash this is discussed in Sec. \ref{sec5c}. 

\renewcommand{\arraystretch}{1.25}
\begin{table}[t]
\caption{Mapping of variables in the method of Lagrange multipliers to the coupled oscillator network}
\label{tab:mapping}
\begin{tabularx}{0.45\textwidth}{||>{\centering\hsize=1.1\hsize}X|>{\centering\arraybackslash\hsize=0.9\hsize}X||}
\hline
\textbf{Problem variable} & \textbf{Physical variable} \\
\hline
\hline
Spin variable $x_i$ & $\left(1/A_\text{sat}\right) \times A_{si}$ \\
\hline
Lagrange multiplier $\lambda_i$ & $4R \times \Lambda_i$ \\
\hline
Coupling matrix $J_{ij}$ & $J_{ij}$ \\
\hline
Lagrange function $L$ & $\lp4R/A^2_{\text{sat}}\rp\times\mathcal{L}$ \\
\hline
Step size $\kappa$ & $1/\left(8RC_{0s}\right)$ \\
\hline
Step size $\kappa'$ & $C_N^2\omega_0^2RA^2_{\text{sat}}/C_{0p}$ \\
\hline
\end{tabularx}
\end{table}

The signal voltages $A_{si}$ of the oscillators play the role of the Ising variables $x_i$, while the $\Lambda_i$ variables play the role of the Lagrange multipliers. The $\Lambda_i$ variables correspond physically to the gain supplied to each oscillator from the pump. In Equation \eqref{eq:circuit_lambda} for $\Lambda_i$, the term $\tfrac{1}{2}C_N\omega_0A_{pi}$ is a negative conductance that corresponds to parametric gain. Since the pump voltage $A_{pi}$ is the only time-varying component of the gain conductance, the time evolution of the Lagrange multipliers $\lambda_i$ is fully contained in the dynamics of the pump oscillator voltages.

Pump depletion performs the role of Lagrange multiplier feedback to constrain the signal voltages. When the system reaches a steady state, all of the signal voltages satisfy the binarization constraint such that $x_i = \pm 1$. This can also be seen in Equation \eqref{eq:dApi_dt}: in steady state ($dA_{pi}/dt = 0$), the amplitude of every oscillator is the same and equals $A_\text{sat}$. Therefore, pump depletion, which is equivalent to the dual step in the Lagrange method, ensures amplitude homogeneity of all the signal voltages in steady state. This new insight supersedes our earlier publication  \cite{vadlamani2020physics}, where we had claimed that a separate feedback circuit would be necessary to implement the $\bm{\lambda}$ feedback. The Lagrange algorithm is entirely self-contained in the dynamics of parametric oscillators.

%terms $\lambda_i=\frac{C_N\omega_0A_{pi}}{2}-\frac{N-1}{4R}$ play the role of the Lagrange multipliers. $\frac{C_N\omega_0A_{pi}}{2}$ may be treated as a negative conductance that supplies gain to the signal. Therefore, the gain conductance of each oscillator (offset by $\frac{N-1}{4R}$) acts as the Lagrange multiplier for that spin. Since the pump voltages $A_{pi}$ are the only time-varying component of the gain conductances, we may say that the pump voltages $A_{pi}$ are the true physical manifestation of the Lagrange multipliers in the circuit. Pump depletion, Eq. \eqref{netpump} performs the role of Lagrange multiplier feedback, Eq. \eqref{}. This new insight supersedes an earlier publication of ours, \cite{}, where we had claimed that a separate feedback circuit would be necessary to implement the $\bm{\lambda}$ feedback. The Lagrange circuit is entirely self-contained.

\subsection{Implementing the Augmented Lagrangian method}
Since the Ising problem does not satisfy strong duality, the Augmented Lagrange function $L_{\alpha}(\bm{x},\bm{\lambda})$ provides a theoretically more optimal solution. For the Ising problem, this is given by:
\begin{equation}
    L_{\alpha}(\bm{x},\bm{\lambda})=L(\bm{x},\bm{\lambda})+\frac{\alpha}{2}\sum_{i=1}^N\lp1-x_i^2\rp^2
\end{equation}
where $L(\bm{x},\bm{\lambda})$ is from Eq. \eqref{eq:IsingLagrange}.

\begin{figure}[t]
\centering
\includegraphics[width=0.48\textwidth]{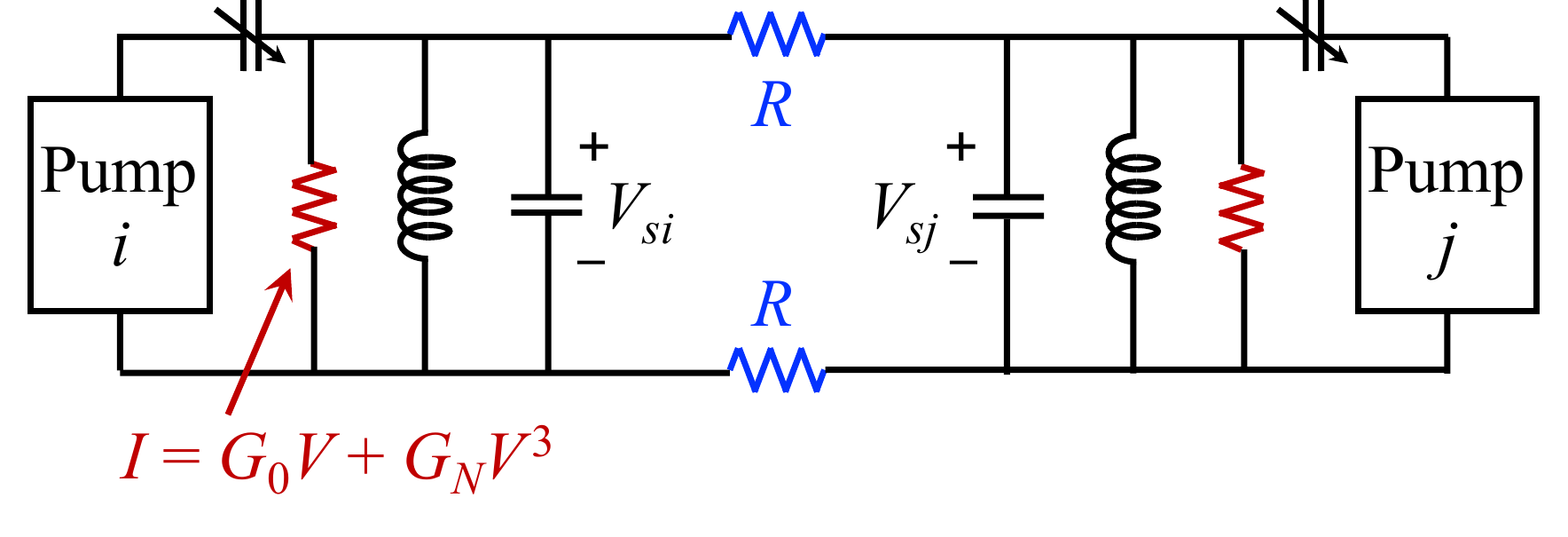}
\caption{Parametric oscillator circuit augmented with a nonlinear resistor to implement the Augmented Lagrangian method, shown for two ferromagnetically coupled spins.}
\label{fig:AugLag}
\end{figure}

The Augmented Lagrange equations of motion are:
\begin{align}
\label{eq:AMMx}
    \frac{dx_i}{dt}&=-2\kappa\left(-\sum_{j=1}^NJ_{ij}x_j-\lambda_ix_i-\alpha x_i+\alpha x_i^3\right)\\
    \frac{d\lambda_i}{dt} &=\kappa'\lp1-x_i^2\rp
\end{align}
The equations are essentially the same as before except for an additional cubic nonlinear term that appears in Eq. \eqref{eq:AMMx}. This nonlinear term, in addition to offering the theoretical optimization advantages discussed in \cite{bertsekas_nonlinear_1999}, also ensures that the signal voltages remain closer to the saturation amplitude than in the plain Lagrange method. To map this term, the parametric oscillator circuit is augmented with a nonlinear resistor in parallel with the signal capacitor with characteristic $I=G_0V+G_NV^3$, as shown in Fig. \ref{fig:AugLag}. In an electrical circuit, a simple practical implementation is a pair of parallel $p$-$n$ junction diodes that conduct in opposite directions.

%\begin{figure}[t]
%\centering
%\includegraphics[scale = 0.42]{PatrickCircuitnonlin.pdf}
%\caption{The coupled LC circuit that implements the Augmented Lagrange method for two %spins. A nonlinear resistor $R_{nlin}$ is placed in parallel with the parametric %capacitor in each spin. The pumping circuit is only represented by an arrow over the %capacitors for brevity.}
%\label{fig:AugLag}
%\end{figure}

Since the additional resistor is in the signal part of the circuit, the pump equations remain unchanged. The equations of motion for the signal circuit, derived in Appendix \ref{appsec4}, are:
\begin{align}
\begin{split}
    \dot{A}_{si}=&\ls-\frac{N_i}{4RC_{0s}} A_{si}+\frac{1}{4RC_{0s}}\sum_{j:j\ne i}J_{ij} A_{sj}\rs+\frac{C_N\omega_0A_{pi}}{2C_{0s}}A_{si}\\
    &-\frac{G_0}{2C_{0s}}A_{si}-\frac{3G_N}{8C_{0s}}A_{si}^3
\end{split}\label{AugLagsigeq}
\end{align}
This equation can be cast into the form of Eq. \eqref{eq:AMMx} by rewriting it as:
\begin{equation}
    \begin{split}
        \dot{A}_{si}=-\frac{1}{4RC_{0s}}&\ls-\sum_{j=1}^NJ_{ij}A_{sj}\right.\\[0.2ex]
        &\left.-4R\lp\Lambda_i-\frac{G_0}{2}-\frac{3G_NA_{\text{sat}}^2}{8}\rp A_{si}\right.\\[0.2ex]
        &\left.-\frac{3G_NRA_{\text{sat}}^2}{2}A_{si}+\frac{3G_NR}{2}A_{si}^3\rs
    \end{split}
\end{equation}
and making the identification:
\begin{align}
    x_i&:=\frac{A_{si}}{A_{\text{sat}}}\\[0.2ex]
    \lambda_i&:=4R\lp\Lambda_i-\frac{G_0}{2}-\frac{3G_NA_{\text{sat}}^2}{8}\rp\\[0.2ex]
    \alpha&:=\frac{3G_NRA^2_{\text{sat}}}{2}
\end{align}

As before, the Lagrange multiplier corresponds to the gains supplied by the pump oscillators, but with an additional fixed offset. Since the nonlinear resistor is in the signal part of the circuit, the pump continues to evolve according to Eq. \eqref{eq:lambdai_lagrange}.

%% I'd rather not advertise this too much because it's a weirder circuit
%We numerically show in the next section that the Augmented Lagrange Multiplier method indeed performs better than the standard method on benchmark problems.

\section{Numerical results}
In this section, we present the results of the numerical simulation of Eqs. \eqref{AugLagsigeq} and \eqref{eq:lambdai_lagrange} for Quadratic Binary optimization problems of sizes 50, 100, 250, and 500 from the BiqMac collection, and MAXCUT problems of sizes 800 and 2000 from the Gset collection. In particular, we worked with the Beasley problems in the BiqMac problem set \cite{wiegele2007biq} and with problems 1-10 (size 800) and problems 22-31 (size 2000) of Gset \cite{gset}. The Beasley binary quadratic problems involve minimizing a quadratic objective function where the feasible set is $0/1$ vectors and the function coefficients are positive and negative integers. Gset problems 1-5 and 22-26 have only 0,1 edge weights while problems 6-10 and 27-31 have -1,0,1 weights. These problems are readily converted to Ising instances through the simple procedure of Appendix \ref{translate}.

\subsection{Parameter choices}
Table \ref{paramvals} lists the circuit parameter definitions and values that were used in the simulations. The prefixes `signal' and `pump' refer to components of the $\omega_0$ and $2\omega_0$ circuits, respectively.

The linear capacitance and inductance values were chosen to set the natural frequency of the signal and pump oscillators to $\omega_0 = 1$GHz and $2\omega_0 = 2$GHz, respectively. The nonlinear capacitance $C_N$ is chosen so that the modulation on the capacitance is $10\%$ of $C_s$ at an applied voltage of 1V. The linear part $C_0$ of the nonlinear capacitance is assumed to be $0$ because any nonzero $C_0$ can be absorbed into $C_s$ and $C_p$ (made clear in the derivations in Appendix \ref{appsec2}). The voltage saturation amplitude $A_{\text{sat}}$ of the signal oscillations is set to 10 mV. 

%To set the coupling resistance value, we define a new quantity\textemdash the average coordination number of the problem\textemdash by computing the cumulative strength (sum) of all connections to every spin and then taking the average. We use this name since, in the special case of 0/1 connections, this quantity is equal to the average number of spins that each spin is connected to.

\renewcommand{\arraystretch}{1.2}
\begin{table}[h]
\caption{Circuit parameter definitions and values} \label{paramvals}
\vspace{1mm}
\begin{tabularx}{0.48\textwidth}{ ||>{\centering\arraybackslash\hsize=1.25\hsize}X|>{\centering\arraybackslash\hsize=0.75\hsize}X|| } 
\hline
\textbf{Parameter} &\textbf{Value}\\
\hline\hline
Signal capacitance ($C_s$) & $1/(2\pi)\ \text{nF}$ \\ 
\hline
Signal inductance ($L_s$) & $1/(2\pi)\ \text{nH}$ \\
\hline
Pump capacitance ($C_p$) & $0.01/(4\pi)\ \text{nF}$ \\
\hline
Pump inductance ($L_p$) & $100/(4\pi)\ \text{nH}$ \\
\hline
Linear connecting cap ($C_0$) &0\\
\hline
Nonlinear connecting cap ($C_N$) & $0.1/(2\pi)\ \text{nF/V}$ \\
\hline
Signal saturation voltage ($A_{\text{sat}}$) & 0.01 V \\
\hline
%Average coordination \# ($\Gamma$) & $\frac{1}{N}\lp\sum_{ij}|J_{ij}|+\frac{1}{2}\sum_{i}|h_i|\rp$\\
%\hline
Common coupling resistance ($R$) & $(500\times \Gamma/47.94)\ \Omega$ \\
\hline
Pump internal resistance ($R_p$) & $\infty$ \\
\hline
Signal internal conductance ($G_s$) & $1/R$\\
\hline
Cubic nonlinear conductance ($G_N$) &$1/(RA_{\text{sat}}^2)$\ $\Omega^{-1}\text{V}^{-2}$\\
\hline
\end{tabularx}
\end{table}

Binary weights $J=\pm1$ can be implemented simply by connecting the signal oscillators with resistors having a common resistance $R$ in the parallel or cross configuration (Fig. \ref{fig:coupling}). Values of $J$ other than $\pm1$ can be constructed using a geometric series of resistances centered at $R$ that encode the binary expansion of $J$ (see Appendix \ref{appsec2b}). We use a different value of $R$ for each problem, set heuristically using the quantity $\Gamma$, which we call the average coordination number of the problem:
\begin{equation}
\Gamma = \frac{1}{N}\lp\sum_{i}\sum_{j:j\neq i}|J_{ij}|+\frac{1}{2}\sum_{i}|h_i|\rp
\end{equation}
In the special case of 0/1 connections, $\Gamma$ is the average number of nonzero connections to each spin. For the first Gset problem of size 800, we empirically found that setting $R=500\,\Omega$ satisfied the slowly-varying amplitude approximation and led to good performance. This problem has an average coordination number of $\Gamma=47.94$. For other problems of average coordination number $\Gamma'$, we set $R=\lp500\Gamma'/\Gamma\rp\Omega$.

In order to ensure that the isomorphism with Lagrange multipliers holds, the pump is assumed to have no internal dissipative loss, unless otherwise noted below. The effect of pump resistance and noise on the performance is discussed later in this section and in Table \ref{tab:performance}. The cubic coefficient $G_N$ of the nonlinear conductance (used to implement the Augmented Lagrangian method) is chosen so that the linear and cubic conductances are equal at the saturation voltage $A_{\text{sat}}$. Further discussion on how varying these parameters affects the solver's performance is provided in Appendix \ref{appsec6}.

%Even in the presence of gain from the pump circuit, this cubic term ensures that signal voltage amplitudes do not stray too far beyond $10$ mV. Further discussion on how varying these parameters affects the solver's performance is provided in the appendix.

%Finally, the cubic nonlinear saturating conductor in the $\omega$ circuit has linear conductance $G_s$ equal to the reciprocal of the common coupling resistance, and the cubic coefficient $G_N$ is chosen so that the linear and cubic conductances become equal at the saturation voltage $A_{\text{sat}}$ of 10$m$V. Even in the presence of gain from the pump circuit, this cubic term ensures that signal voltage amplitudes do not stray too far beyond $10m$V. Further discussion on how varying these parameters affects the solver's performance is provided in the appendix.

% \subsection{Results}

\subsection{Dynamics of the solver}
\begin{figure}[t]
\centering
\includegraphics[width=0.48\textwidth]{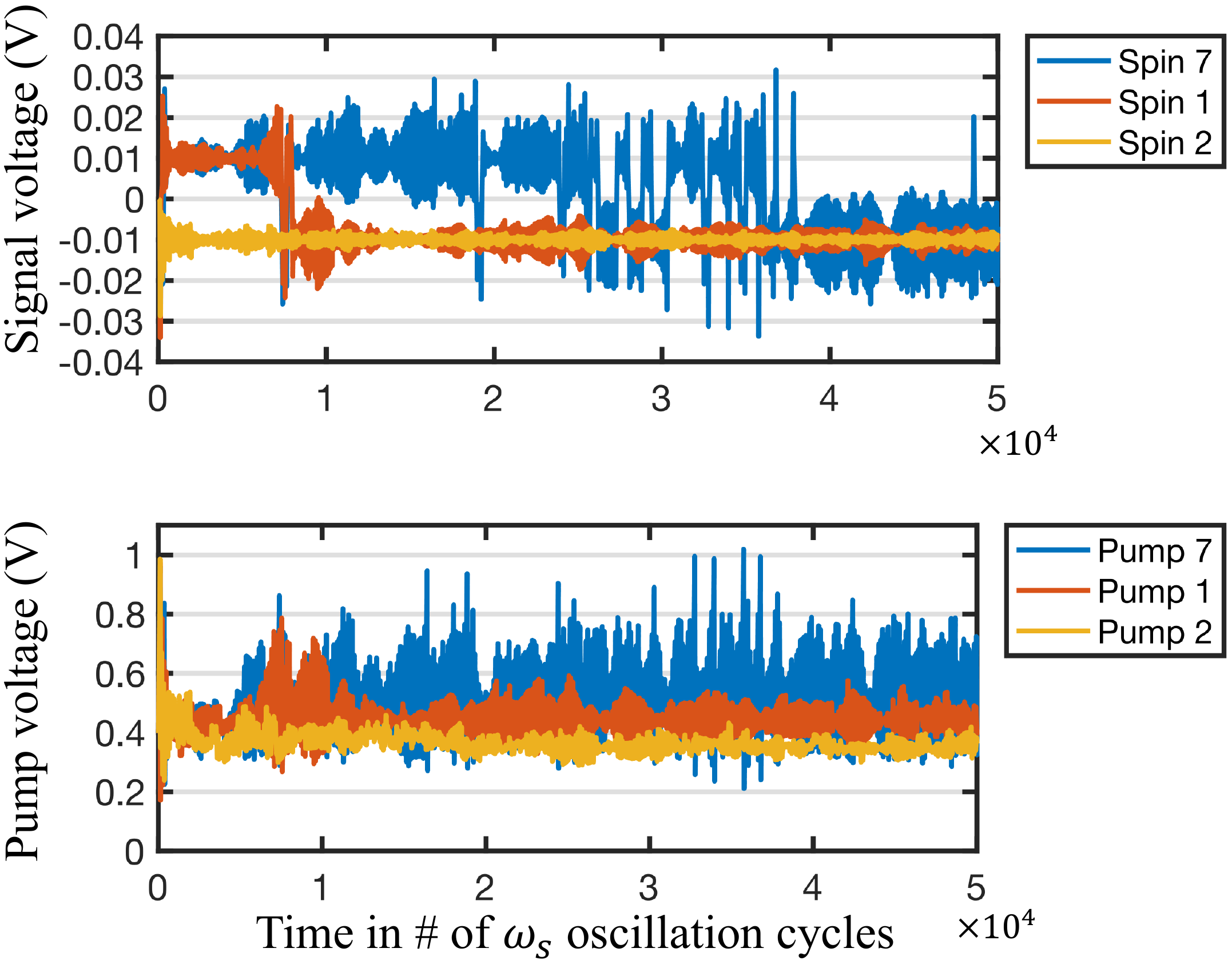}
\caption{Time evolution of the signal (top) and pump (bottom) oscillator voltages for spins 1, 2, 7 in an 800-variable Gset problem. The signal voltages start from noise before eventually saturating to $\pm$10 mV.}
\label{fig:voltvstime}
\end{figure}

The slowly varying amplitude equations \eqref{AugLagsigeq} and \eqref{eq:lambdai_lagrange} were simulated using MATLAB's built-in \texttt{ode45} ODE solver for a total time of $50\,\mu$s. All the signal capacitor voltages start at the noise level $\sqrt{\frac{kT}{C_s}}\approx 5\mu V$ while the initial pump voltages are set such that there is gain right from $t=0$. Further details of the initial conditions and the simulation setup are provided in Appendix \ref{appsec6}. The signal and pump oscillator voltages for the first 800-vertex problem in Gset are shown in Fig. \ref{fig:voltvstime}. The oscillators corresponding to spins 1, 2, and 7 are plotted to depict the diversity of behaviors observed in the system: Spin 2 starts out near the noise level but immediately settles down to a steady state of -10mV (logical -1), Spin 1 flips from logical +1 to $-$1 after an initial period of evolution, and Spin 7 undergoes rapid repeated flipping between -1 and +1 and has relatively large fluctuations in its oscillation amplitude.

The time evolution of the pump voltages is shown in Fig. \ref{fig:voltvstime} (bottom). The pump voltage indicates how much parametric gain is being supplied to the corresponding signal oscillator in order to maintain a steady-state amplitude of $\pm$10 mV. As explained previously, the pump voltage dynamics directly tracks the time evolution of the Lagrange multipliers. Spin 7, which has large fluctuations in the signal voltage and thus frequent deviations from the binary constraint, has correspondingly large fluctuations in its pump voltage.

\begin{figure}[t]
\centering
\includegraphics[width=0.47\textwidth]{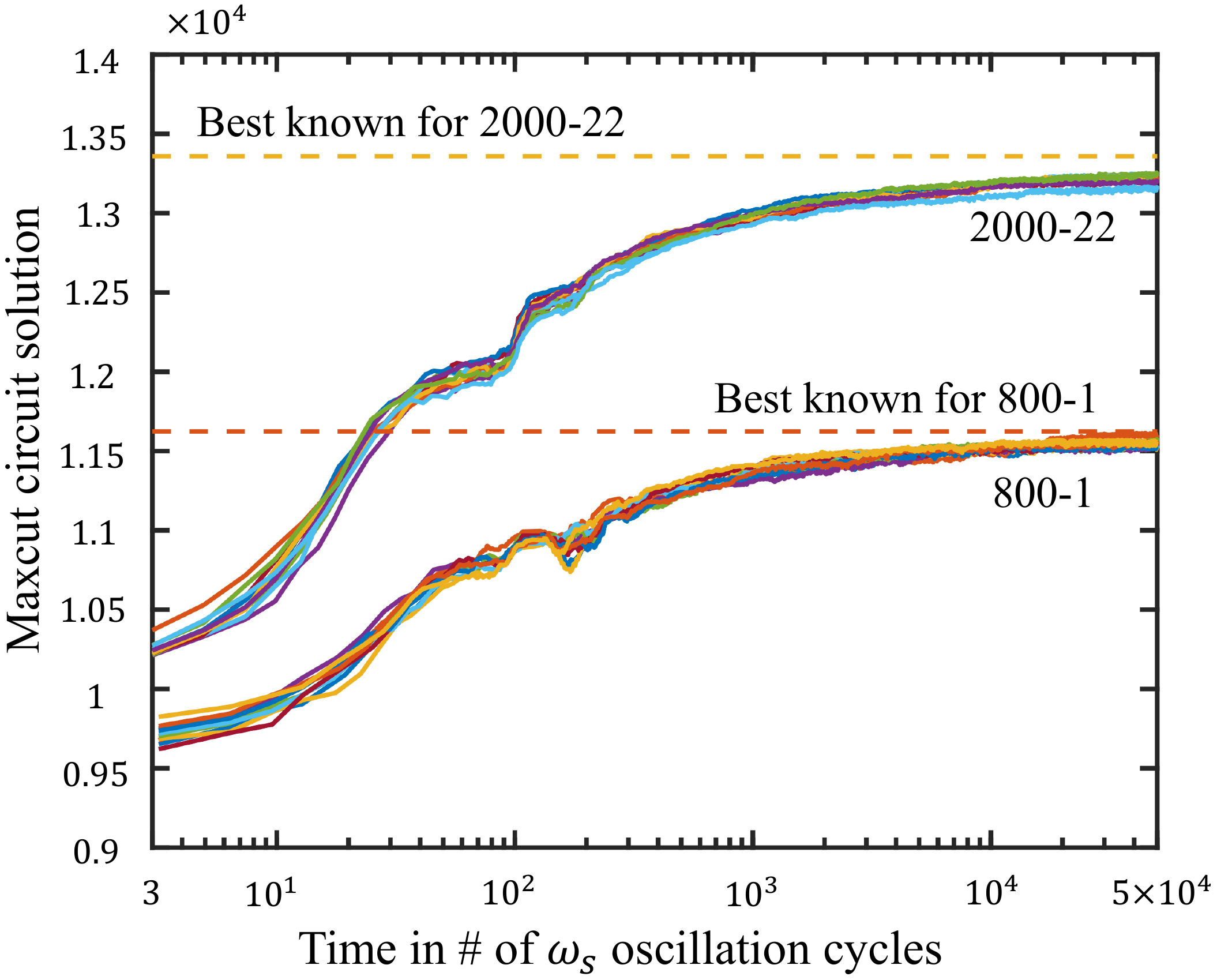}
\caption{Time evolution of the objective function as the circuit evolves. Different colors correspond to 10 independent, randomized runs of the circuit starting from noise for each of two problems: Gset 1 (800 vertices) and 22 (2000 vertices).}
\label{fig:maxcutvstime}
\end{figure}

At each point in time, the collection of signal voltages $A_{si}$ can be converted to a binary solution vector by taking the sign of each element. This allows the computation of an instantaneous MaxCut value, shown in Fig. \ref{fig:maxcutvstime} for two problems: Gset \#1 with 800 variables and Gset \#22 with 2000 variables. Most of the progress toward the optimum is made at early times, with a slowdown in improvements as time progresses. The best instantaneous objective value within the 50 $\mu$s simulation window is declared as the solution of the run.

\subsection{Quality of solution}\label{sec5c}
To understand how performance scales with size, we used BiqMac benchmark problems of size 50, 100, 250, and 500, and Gset benchmark problems of size 800 and 2000. Our problem set consisted of 10 problems of each size for a total of 60 problems. The solver was run 10 times with random independent initial conditions on each problem and the best and median solutions obtained over the 10 runs were recorded for each problem. The results for Gset problems 1, 2 (800 spins, 0,1 weights) and problems 6, 7 (800 spins, -1,0,1 weights) are presented in Table \ref{tab:performance}. A more comprehensive list is provided in Appendix \ref{appsec6b}. 

\renewcommand{\arraystretch}{1.15}
\begin{table*}
\caption{Performance on selected Gset 800-variable MaxCut problems of our approach, compared to other known algorithms. The best and median solution quality are reported for 10 independent runs for the last five columns, while the results for Leleu are for 20 runs \cite{leleu_destabilization_2019}. The pump quality factor $Q_p$ parameterizes the internal loss in the pump oscillator.}
%Note that the ratio of the Goemans-Williamson solution and the impied upper bound is not $\approx88\%$ in the last two rows because those problems have negative weight edges.}
\label{tab:performance}
\vspace{1mm}
\begin{tabularx}{\textwidth}{||
>{\centering\hsize=0.75\hsize}X|
>{\centering\hsize=1\hsize}X|
>{\centering\hsize=1\hsize}X|
>{\centering\hsize=1\hsize}X|
>{\centering\hsize=1.125\hsize}X|
>{\centering\hsize=1.125\hsize}X|
>{\centering\hsize=1\hsize}X|
>{\centering\hsize=1\hsize}X|
>{\centering\arraybackslash\hsize=1\hsize}X|| } 
\hline
\multirow{3}{*}{Problem} & \multirow{3}{*}{\shortstack{Goemans-\\Williamson}} &  \multirow{3}{*}{Metric} & \multirow{3}{*}{\shortstack{Leleu \textit{et al.}\\\cite{leleu_destabilization_2019}}} & \multirow{3}{*}{\shortstack{Oscillators,\\Plain\\Lagrange}} & \multirow{3}{*}{\shortstack{Oscillators,\\Augmented\\Lagrange}} & \multicolumn{3}{c||}{Oscillators, Augmented Lagrange} \\
  & & & & & & \multicolumn{3}{c||}{with thermal noise} \\
\cline{7-9}
  & & & & & &$Q_p=100$ &$Q_p=500$ & $Q_p=5000$\\
\hline\hline
 \multirow{2}{*}{1} & 11272 &best &11624 & 11580 &11613 &9963 &11532 &11592  \\
 & UB: 12838 &median &11624 &11552 &11558 &9941 &11512 &11578\\
 \hline
 \multirow{2}{*}{2} & 11277 &best &11620 & 11575 &11596 &9941 &11531 &11604\\
  & UB: 12844 &median &11620 &11554 &11572 &9933 &11505 &11584\\
 \hline
 \multirow{2}{*}{6} & 1813 &best &2178  & 2143 &2173 &470 &2088 &2162\\
  & UB: 3387 &median &2178 &2124 &2144 &439 &2076 &2136\\
 \hline
 \multirow{2}{*}{7} & 1652 &best &2006 & 1975 &1973 &327 &1922 &1990\\
  & UB: 3224 &median &2006 &1950 &1955 &274 &1904 &1967\\
\hline
\end{tabularx}
\end{table*}

We use the performance of the well-known Goemans-Williamson algorithm as a baseline for comparison, as well as to provide a theoretical upper bound (UB) on the MaxCut solution quality. The best known solutions to these specific MaxCut problem instances are from Leleu \textit{et al.} \cite{leleu_destabilization_2019}. For our coupled oscillator approach, we include the quality of the solution found without and with the nonlinear resistor, i.e. for the plain Lagrange multipliers and the Augmented Lagrange methods, respectively. Finally, we include the results of the coupled oscillator network under less ideal conditions: the pump circuit is made lossy (parameterized by the quality factor of the pump oscillator, $Q_p=R_p\sqrt{C_p/L_p}$), and Johnson thermal noise is incorporated into both the signal and pump circuits. The noise model is described in Appendix \ref{appsec2c}.

%In Table \ref{tab:performance}, we list for each problem the performance of the Goemans-Williamson algorithm, the upper bound (UB) implied by it, and the best and median solutions obtained by Leleu et al.'s algorithm \cite{leleu_destabilization_2019}, our coupled oscillator approach without nonlinear resistive saturation (plain Lagrange Multipliers) and with it (Augmented Lagrange Multipliers), and the results when the pump circuit is made lossy and Johnson noise is incorporated into both the signal and pump circuits. The exact noise model is described in the appendix. A few key takeaways from the table are:

We note several key findings from these results. First, the coupled parametric oscillator network far outperforms the basic Goemans-Williamson algorithm. Secondly, the physical system that implements the Augmented Lagrange method generally performs better than the plain Lagrange method, though the difference between the two methods is not always significant.

% Please expand further on why this is not surprising
% Also am I right to conclude noise doesn't do much, mostly it's the loss? If so, please state that
Introducing loss in the pump circuit leads to a reduction in performance. This is not surprising because the addition of pump loss breaks the exact correspondence with Lagrange multipliers as pointed out in Section \ref{sec:correspondence}. The performance deterioration increases as the pump quality factor is reduced, with the results for $Q_p > 5000$, even with thermal noise included in both the signal and pump circuits, being similar to the lossless, noiseless case.

Finally, the algorithm in Leleu \textit{et al} \cite{leleu_destabilization_2019} finds higher-quality solutions compared to the Lagrange multiplier solver. This is possibly due to non-gradient chaotic dynamics that does not get stuck at fixed points or limit cycles. Lagrange multipliers on the other hand follow gradient-based dynamics in the form of alternating descent and ascent. Unlike the method of Lagrange multipliers, the algorithm of Leleu \textit{et al} is not known to have a direct mapping to a physical system.

The remainder of the oscillator results in this section (and the appendix) are for the Augmented Lagrange method, assuming a lossless pump and no noise.

% \begin{enumerate}
%     \item The Augmented Lagrange method performs better than the plain version.
%     \item Introducing loss in the pump circuit leads to a reduction in performance\textemdash this is not surprising because the addition of loss breaks the exact correspondence with Lagrange multipliers as pointed out in Section \ref{}. Further, the performance deterioration increases as the pump quality factor is reduced with the results for high Q being similar to the lossless, noiseless case.
%     \item Leleu's method outperforms Lagrange multipliers, possibly due to non-gradient chaotic dynamics that does not get stuck at fixed points or limit cycles. Lagrange multipliers on the other hand follows gradient-based dynamics in the form of alternating descent and ascent.
% \end{enumerate}

\begin{figure}[t]
\centering
\includegraphics[width=0.47\textwidth]{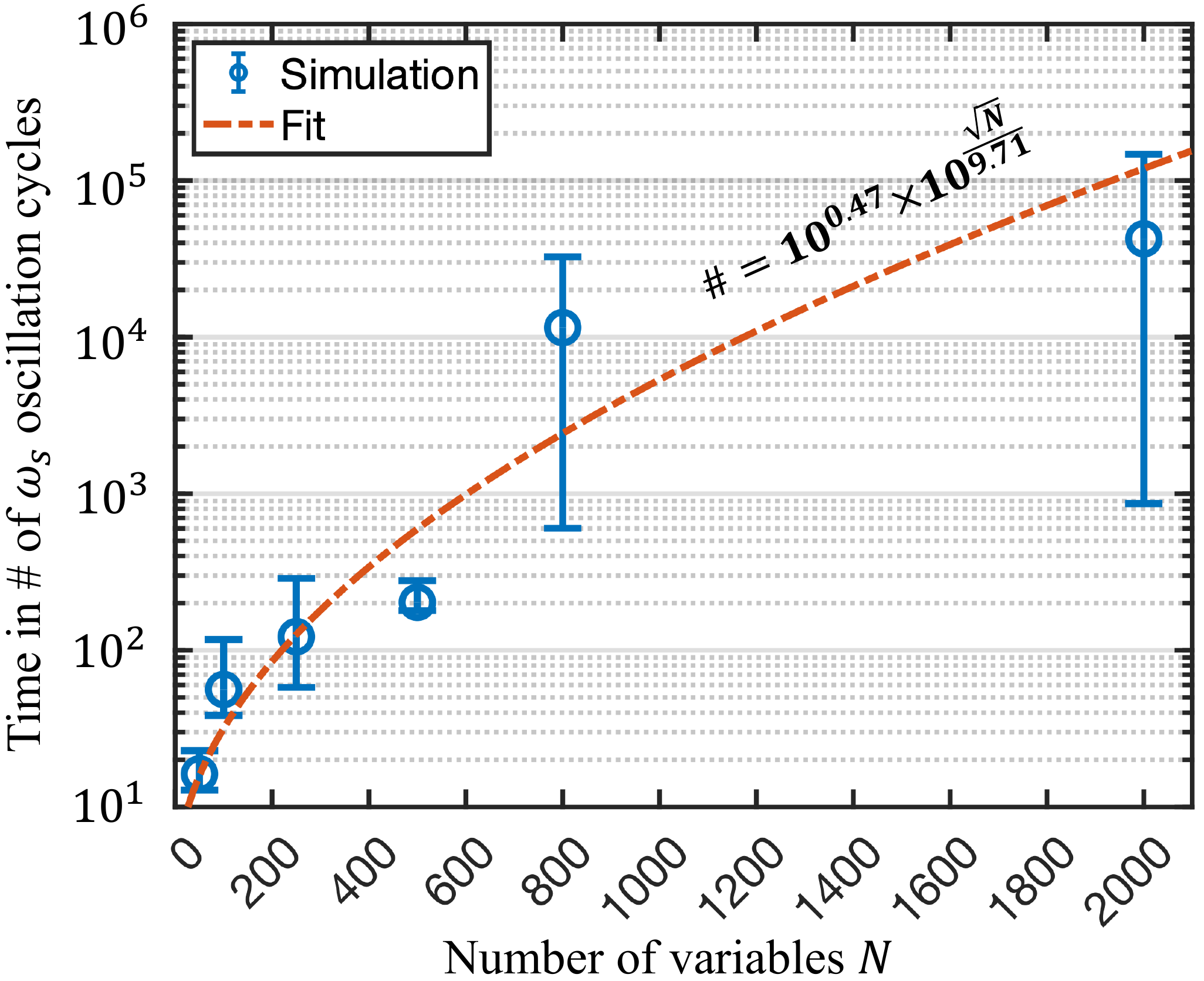}
\caption{TTS as a function of problem size. Ten problems were solved per problem size, and the median, 25$^\text{th}$, and 75$^\text{th}$ percentile for the TTS are shown. The first four points, for sizes 50, 100, 250, and 500, are for the BiqMac problems while the latter two points, for sizes 800 and 2000, are for Gset. The number of cycles scales as $10^{\sqrt{N}/9.71}$.}
\label{fig:timetosol}
\end{figure}

\subsection{Time-to-solution}
Next, we extract the dependence of the time-to-solution (TTS) on the problem size. A run of the solver on a given problem is considered \textit{successful} if the instantaneous objective function value breaches 97\% of the best-known value for that problem at some point during the 50$\mu s$ duration of the run. We define the TTS for a successful run as the first time the 97\% mark is crossed. For an unsuccessful run, the TTS is the full 50$\mu s$. The TTS for the problem is then equal to the sum of the TTS of all the runs divided by the number of successful runs. This metric measures the average time spent between two successes. Fig. \ref{fig:timetosol} shows how the TTS depends on the number of variables $N$ in our problem set. Though these problems are drawn from two different benchmark sets, the TTS, in number of $\omega_0$ oscillation cycles, scales as $10^{0.47}\times10^{\frac{\sqrt{N}}{9.71}}=10^{0.47}\times1.27^{\sqrt{N}}$ which is $\mathcal{O}(2^{\sqrt{N}})$, corroborating previous work on solvers of this type that noted similar scaling \cite{hamerly2019experimental,patel2022logically}. 

\subsection{Robustness to coupling resistance imperfections}

\begin{figure}[t]
\centering
\includegraphics[width=0.47\textwidth]{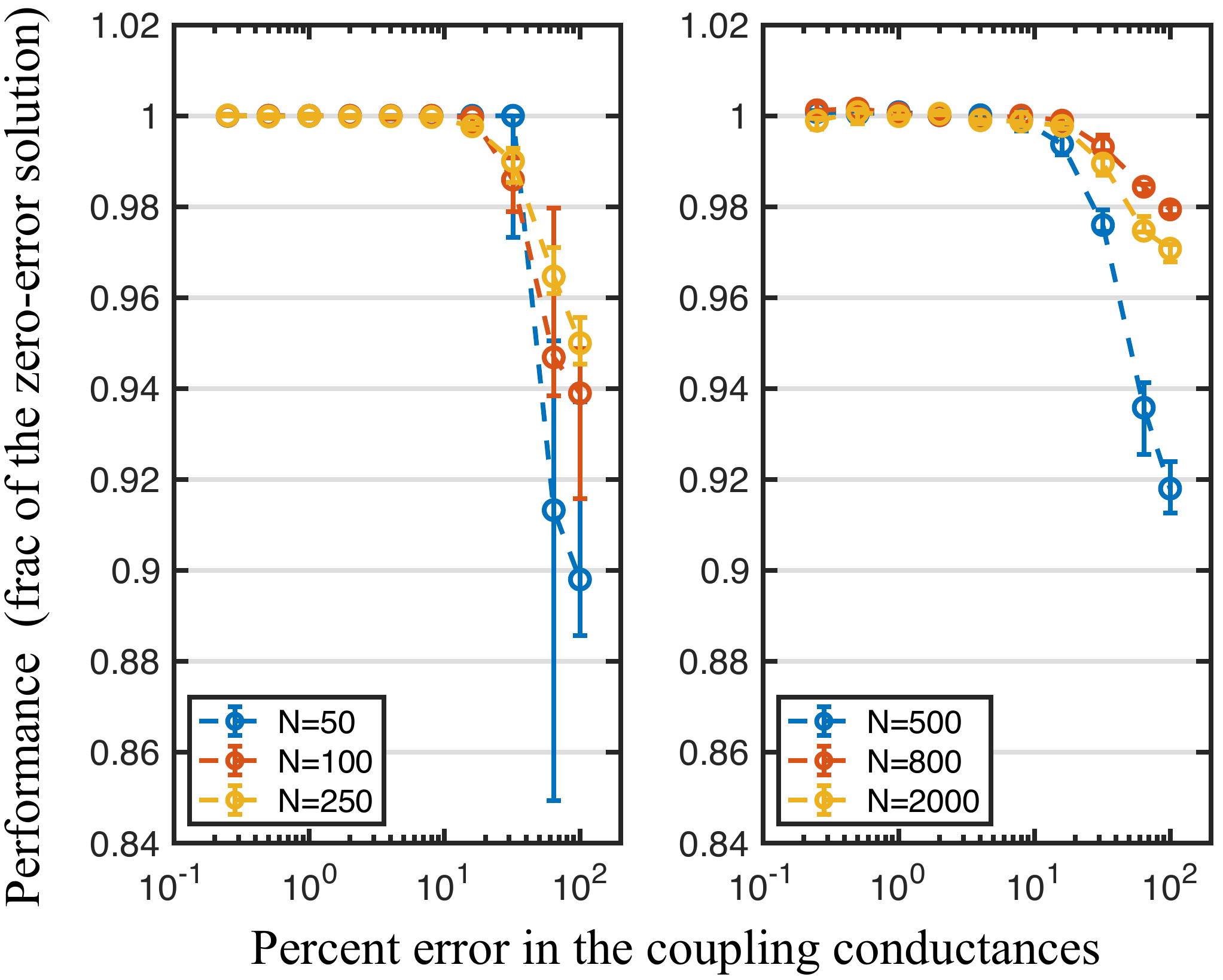}
\caption{Effect of error in the coupling conductance values on solution quality, shown for six problem sizes. The performance is normalized to the solution quality with zero conductance errors.}
\label{fig:conductance_errors}
\end{figure}

Sensitivity to component imperfections is one of the long-standing criticisms of analog computers. The present application has some built-in tolerance to these imperfections because of the fact that the problem demands binary answers, even though the processing is done on analog signals in continuous time. In our network of coupled parametric oscillators, a main source of component imperfections is the (up to) $N^2$ resistors connecting the oscillators together. A connection weight $J_{ij}$ is proportional to the conductance of the connecting resistors, given by $G_{ij}=|J_{ij}|/R$. Errors in these conductance values can cause the wrong problem to be solved by the hardware, in turn leading to non-optimal solutions to the original problem.

We find that in general, the Lagrange multiplier oscillator network is relatively insensitive to errors in the coupling resistors, as shown in Fig. \ref{fig:conductance_errors}. We assumed that the conductance $G_{ij}$ had a Gaussian distribution with a mean given by $|J_{ij}|/R$ and a standard deviation that is a certain percentage of the mean. For each problem size and error value, one specific problem was chosen and solutions were sampled from 10 circuits with randomized conductance errors. Fig. \ref{fig:conductance_errors} indicates that conductance errors as large as 10\% are tolerated without significant reduction in quality of solution even for 2000-spin problems. This level of precision is well within the capability of modern programmble resistive memory devices \cite{xiao2020analog}.

\section{Conclusion}
In this paper, we studied the dynamics of coupled parametric oscillator Ising solvers and showed that the system exactly performs Lagrange multiplier primal-dual optimization. The signal oscillator voltages represent the binary problem variables while the pump oscillator voltages represent the corresponding Lagrange multipliers. The equations of motion of the signal and pump implement the alternating primal (descent) and dual (ascent) equations of the Lagrange multiplier method. A more sophisticated algorithm, the Augmented Lagrange Multiplier method, can be implemented by introducing appropriate nonlinear saturating resistors into the circuit. The simulated numerical performance of the method is competitive with the current best known heuristic Ising solver proposed by Leleu et al. \cite{leleu_destabilization_2019}. In the future, it may be possible to augment the oscillator system with the chaotic dynamics of the Leleu solver to develop powerful new discrete optimization algorithms that can also be implemented by dynamical physical systems.

We showed numerically that the time-to-solution scaled as $\mathcal{O}(2^{\sqrt{N}})$ where $N$ is the problem size\textemdash a result that is consistent with other work in the literature \cite{hamerly2019experimental,patel2022logically}. We also showed that the quality of solutions obtained by the oscillator solvers was robust to errors in the circuit components used to program in the $J_{ij}$. This encouraging result suggests a promising research direction where such circuit solvers are designed for a multitude of important optimization problems and are used as analog co-processors or accelerators alongside standard digital chips. We hope this work will instigate further research into the design of physical systems that naturally perform optimization (physical optimizers) of various flavors for important applications like machine learning.

\begin{acknowledgments}
We gratefully acknowledge useful discussions with Dr. Ryan Hamerly. The work of S.K.V., T.P.X., and E.Y. was supported by the NSF through the Center for Energy Efficient Electronics Science (E\textsuperscript{3}S) under Award ECCS-0939514 and the Office of Naval Research under Grant N00014-14-1-0505.
\end{acknowledgments}

% \appendix

\nocite{*}

\bibliography{apssamp}% Produces the bibliography via BibTeX.
% \section{Appendix}
\appendix

\begin{widetext}
\section{Single parametric LC oscillator\textemdash equations of motion}\label{appsec1}

\begin{figure}[h]
\centering
\includegraphics[width=0.55\textwidth]{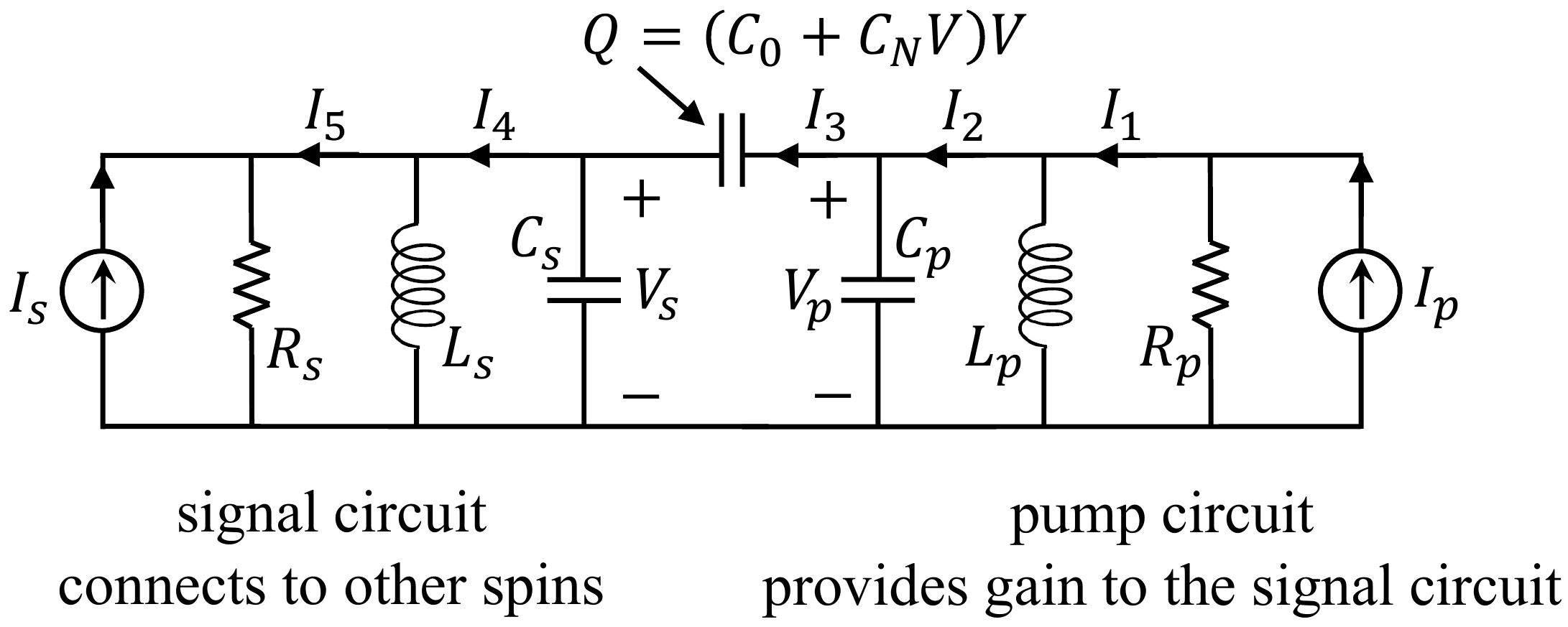}
\caption{The signal LC oscillator on the left and the pump LC oscillator on the right are connected by a nonlinear capacitor.}
\label{fig:apppumpckt}
\end{figure}

Before we start the derivation, we note that the current source in the signal circuit, $I_s$, is just a noise source in our system. Therefore, it can be dropped while considering the evolution of the system. We retain it in the current derivation simply to obtain general expressions but shall drop it as soon as the discussion specializes to our situation. 

The circuit equations for Fig. \ref{fig:apppumpckt} are:
\begin{align}
C_s\dot{V}_s=I_3-I_4,&\ \ C_p\dot{V}_p=I_2-I_3\label{suppeq1}\\
V_s=L_s\lp\dot{I}_4-\dot{I}_5\rp,&\ \ V_p=L_p\lp\dot{I}_1-\dot{I}_2\rp\label{suppeq2}\\
V_s=R_s\lp I_s+I_5\rp,&\ \ V_p=R_p\lp I_p-I_1\rp\label{suppeq3}\\
I_3=C_0\lp \dot{V}_p-\dot{V}_s\rp+2C_N&\lp V_p-V_s\rp\lp \dot{V}_p-\dot{V}_s\rp\label{suppeq4}
\end{align}

$I_3$ can be eliminated by substituting Eq. \eqref{suppeq4} into Eq. \eqref{suppeq1}. Then, Eqs. \eqref{suppeq1} and \eqref{suppeq3} can be used to express $I_1$, $I_2$, $I_4$, and $I_5$ in terms of voltages and the current sources. Finally, plugging all these expressions into Eqs. \eqref{suppeq2} yields:
% \begin{widetext}
    \begin{align}
        V_s&=L_s\lp C_0\lp\ddot{V}_p-\ddot{V}_s\rp+2C_N\lp\dot{V}_p-\dot{V}_s\rp^2+2C_N\lp V_p-V_s\rp\lp\ddot{V}_p-\ddot{V}_s\rp-C_s\ddot{V}_s-\frac{\dot{V}_s}{R_s}+\dot{I}_s\rp\label{suppsinglesignal}\\
        V_p&=L_p\lp -C_0\lp\ddot{V}_p-\ddot{V}_s\rp-2C_N\lp\dot{V}_p-\dot{V}_s\rp^2-2C_N\lp V_p-V_s\rp\lp\ddot{V}_p-\ddot{V}_s\rp-C_p\ddot{V}_p-\frac{\dot{V}_p}{R_p}+\dot{I}_p\rp\label{suppsinglepump}
    \end{align}

In Eq. \eqref{suppsinglesignal}, we retain only terms that oscillate at $\omega_0$ or contribute to oscillations at $\omega_0$. Similarly, in Eq. \eqref{suppsinglepump} we retain only terms that oscillate at $2\omega_0$ or contribute to oscillations at $2\omega_0$. These equations simplify to:
\begin{align}
    \dot{I}_s=&\frac{V_s}{L_s}+\lp C_0+C_s\rp\ddot{V}_s+\frac{\dot{V}_s}{R_s}+4C_N\dot{V}_p\dot{V}_s+2C_NV_p\ddot{V}_s+2C_NV_s\ddot{V}_p\label{suppIsred}\\
    \dot{I}_p=&\frac{V_p}{L_p}+\lp C_0+C_p\rp\ddot{V}_p+\frac{\dot{V}_p}{R_p}+2C_N\dot{V}_s^2+2C_NV_s\ddot{V}_s\label{suppIpred}
\end{align}

At this point, we make the redefinition $C_s:=C_0+C_s$ and $C_p:=C_0+C_p$ for notational convenience. Next, we perform the slowly-varying amplitude approximation by expressing all the currents and voltages involved as follows: 

    \begin{alignat}{2}
        &I_p=\frac{I_{pc}}{2}e^{2i\omega_0 t}+\text{c.c}\ \ \ &&\dot{I}_p=2i\omega_0\frac{I_{pc}}{2}e^{2i\omega_0 t}+\text{c.c}\\
        &I_s=\frac{I_{sc}-iI_{ss}}{2}e^{i\lp\omega_0 t+\phi_s\rp}+\text{c.c}\ \ \ &&\dot{I}_s=i\omega_0\frac{I_{sc}-iI_{ss}}{2}e^{i\lp\omega_0 t+\phi_s\rp}+\text{c.c}\\
        &V_s=\frac{A_s-iB_s}{2}e^{i\lp\omega_0 t+\phi_s\rp}+\text{c.c}\ \ \ &&\dot{V}_s=i\omega_0\frac{A_s-iB_s}{2}e^{i\lp\omega_0 t+\phi_s\rp}+\frac{\dot{A}_s-i\dot{B}_s}{2}e^{i\lp\omega_0 t+\phi_s\rp}+\text{c.c}\\
        &\ddot{V}_s=2i\omega_0\frac{\dot{A}_s-i\dot{B}_s}{2}e^{i\lp\omega_0 t+\phi_s\rp}-\omega_0^2\frac{A_s-iB_s}{2}e^{i\lp\omega_0 t+\phi_s\rp}+\text{c.c}\\
        &V_p=\frac{A_p}{2}e^{i\lp2\omega_0 t\rp}+\text{c.c}\ &&\dot{V}_p=2i\omega_0\frac{A_p}{2}e^{i\lp2\omega_0 t\rp}+\frac{\dot{A}_p}{2}e^{i\lp2\omega_0 t\rp}+\text{c.c}\\
        &\ddot{V}_p=4i\omega_0\frac{\dot{A}_p}{2}e^{i\lp2\omega_0 t\rp}-4\omega_0^2\frac{A_p}{2}e^{i\lp2\omega_0 t\rp}+\text{c.c}
    \end{alignat}
where $A_s$ is the cosine component of $V_s$ and $B_s$ is its sine component. Plugging these expressions into Eq \eqref{suppIsred}, we get:
\begin{equation}
\begin{split}
    i\omega_0 \lp I_{sc}-iI_{ss}\rp=&\frac{A_s-iB_s}{L_s}+C_s\lp2i\omega_0\lp\dot{A}_s-i\dot{B}_s\rp-\omega_0^2\lp A_s-iB_s\rp\rp+\frac{i\omega_0\lp A_s-iB_s\rp+\dot{A}_s-i\dot{B}_s}{R_s}\\
    &+C_N\ls2\lp2i\omega_0A_p+\dot{A}_p\rp\lp-i\omega_0\lp A_s+iB_s\rp+\dot{A}_s+i\dot{B}_s\rp\right.\\
    &\left.+A_p\lp-2i\omega_0\lp\dot{A}_s+i\dot{B}_s\rp-\omega_0^2\lp A_s+iB_s\rp\rp+\lp A_s+iB_s\rp\lp4i\omega_0\dot{A}_p-4\omega_0^2A_p\rp\rs e^{i\lp-2\phi_s\rp}
\end{split}\label{signallongslowly}
\end{equation}
\end{widetext}
Equating the imaginary parts on both sides, recognizing that $1/\omega_0^2=L_1C_s$, rearranging terms, and setting $\phi_s=3\pi/4$, we have:
\begin{equation}
\begin{split}
    \dot{A}_s=&\frac{I_{sc}}{2C_s}-\frac{A_s}{2R_sC_s}+\frac{C_N\omega_0A_p}{2C_s}A_s+\frac{\dot{B}_s}{2R_s\omega_0C_s}\\
    &-\frac{C_N}{2\omega_0C_s}\ls2\dot{A}_s\dot{A}_p-2\omega_0\dot{A}_pB_s-2\omega_0\dot{B}_sA_p\rs
\end{split}\label{asbig}
\end{equation}
The first term of the first line on the right hand side is the injection from the current source, the second term is the internal resistive loss, the third is the gain provided by the pump to the signal cosine component. The fourth term can be ignored because its magnitude is $\mathcal{O}\lp \frac{\text{loss coeff.}}{\omega_0}\rp$ which is small by the slowly-varying amplitude approximation. The first term in the square brackets on the last line is small compared to the third term of the first line (again by the slowly varying approximation) and can be dropped. Finally, the second and third terms in the square brackets of the last line can be dropped too because they are of size $\mathcal{O}\lp \frac{\text{gain coeff.}}{\omega_0}\rp$. The cosine amplitude dynamics is then:
\begin{equation}
    \dot{A}_s=\frac{I_{sc}}{2C_s}-\frac{A_s}{2R_sC_s}+\frac{C_N\omega_0A_p}{2C_s}A_s
\end{equation}
Equating the real parts on both sides of Eq. \eqref{signallongslowly}, we get for the amplitude of the sine component:
\begin{equation}
\begin{split}
    \dot{B}_s=&\frac{I_{ss}}{2C_s}-\frac{B_s}{2R_sC_s}-\frac{C_N\omega_0A_p}{2C_s}B_s-\frac{\dot{A}_s}{2R_s\omega_0C_s}\\
    &+\frac{C_N}{2\omega_0C_s}\ls2\dot{B}_s\dot{A}_p+2\omega_0\dot{A}_pA_s+2\omega_0\dot{A}_sA_p\rs
\end{split}\label{bsbig}
\end{equation}
The third term on the first line is a parametric \emph{loss} term and not a gain term due to which the sine component never grows to the same order of magnitude as the cosine component\textemdash this shall be verified in a bit.

\begin{figure}[h]
\centering
\includegraphics[width=0.47\textwidth]{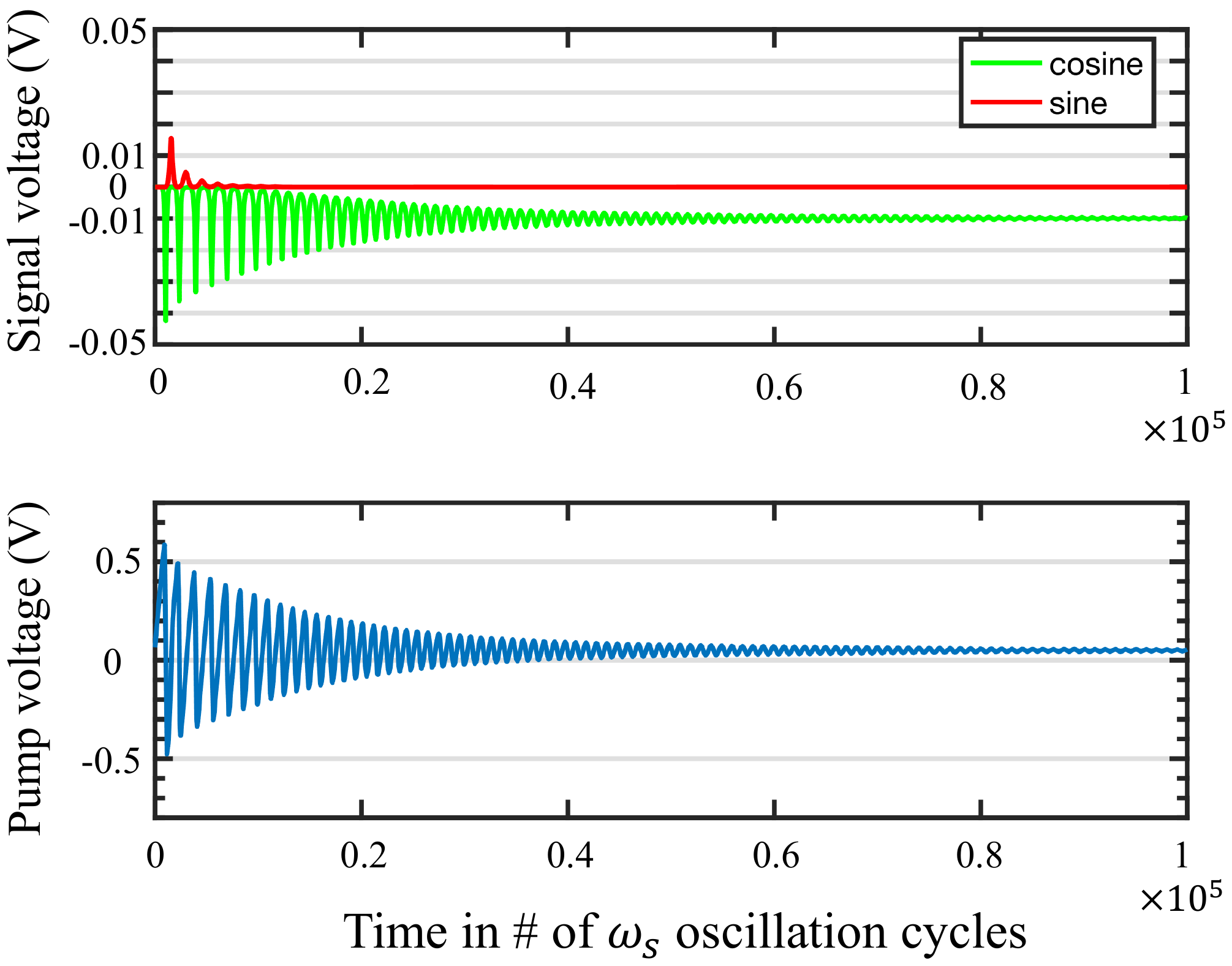}
\caption{Simulation of the slowly varying amplitude equations for a single spin system\textemdash a signal oscillator connected to a pump oscillator. The sine component decays to 0 early in the evolution.}\label{onespin}
\end{figure}

The equivalent of Eqs. \eqref{asbig} and \eqref{bsbig} for the pump circuit is:
\begin{equation}
\begin{split}
    \dot{A_p} =& \frac{I_{pc}}{2C_p}-\frac{A_p}{2R_pC_p}+\frac{C_N}{4C_p\omega_0}\lp \dot{A}_s^2-\dot{B}_s^2\rp\\
    &+\frac{C_N}{4C_p\omega_0}\lp2\omega_0^2\lp B_s^2-A_s^2\rp+4\omega_0\lp A_s\dot{B}_s+\dot{A}_sB_s\rp\rp
\end{split}\label{apbig}
\end{equation}
Simulating Eqs. \eqref{asbig}, \eqref{bsbig}, and \eqref{apbig} using MATLAB's \texttt{ode15i} implicit ODE solver leads to the plots in Fig. \ref{onespin}, confirming that the sine component $B_s$ decays to 0 very early and can be ignored. It shall henceforth be dropped in all our equations. 
Performing the slowly varying amplitude approximation on \eqref{apbig} and dropping terms that contain $B_s$, we obtain the following final pump amplitude evolution equation:
\begin{equation}
    \dot{A}_p=\frac{I_{pc}}{2C_p}-\frac{A_p}{2R_pC_p}-\frac{C_N\omega_0A_s^2}{2C_p}
\end{equation}

\section{Coupled parametric LC oscillators\textemdash equations of motion}\label{appsec2}
In this section, we derive the equations of motion for a network of $N$ coupled parametric oscillators with all-to-all coupling with $J_{ij}$ taking values $\pm 1$. The general case of sparser/non $\pm1$ coupling is dealt with later on in this section.
\begin{widetext}

\begin{figure}[h]
\centering
\includegraphics[width=0.7\textwidth]{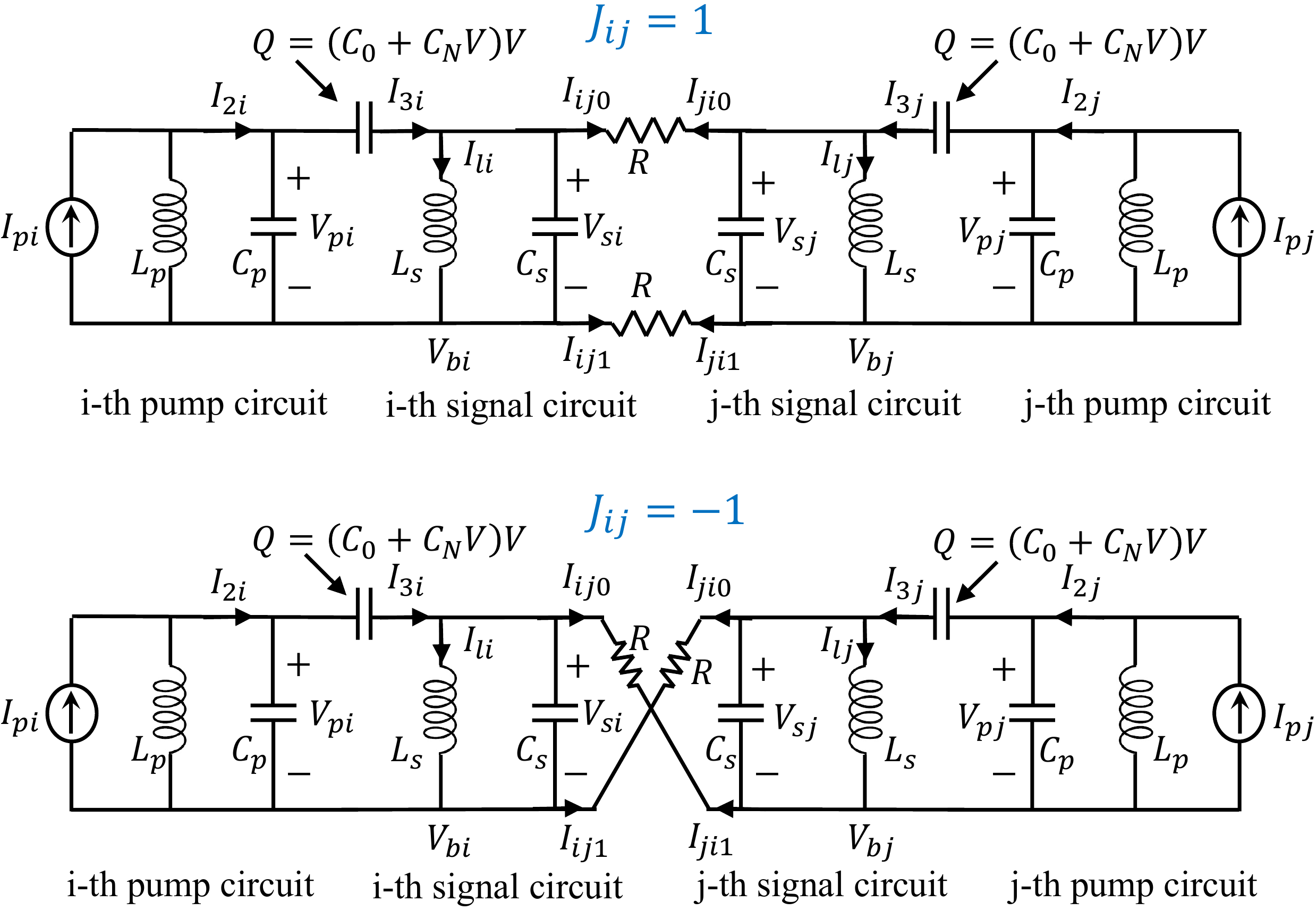}
\caption{The various currents and voltages when one focuses on the coupling between the $i$-th and $j$-th parametric oscillators. The top and bottom panels show the $J_{ij}=1$ and $J_{ij}=-1$ cases respectively.}\label{twospin}
\end{figure}

\end{widetext}

Let the parametric oscillators be labelled from $i=1$ to $i=N$. The notation we will use is indicated in Fig. \ref{twospin} where we focus on the coupling between the $i$-th and $j$-th parametric oscillators. One of the terminals of the capacitor in the oscillator labelled $i=1$ (not shown in the figure) is arbitrarily chosen as its `bottom' terminal, and its other terminal is labelled its `top' terminal. For each oscillator that is connected to $i=1$ through a +1 connection, the terminal in that oscillator that is directly connected to the bottom terminal of $i=1$ is labelled its `bottom' terminal. Similarly, for each oscillator that is connected to $i=1$ through a -1 connection, the terminal in that oscillator that is directly connected to the bottom terminal of $i=1$ is labelled its \emph{top terminal}. We continue this process until all terminals in the circuit get labelled. If two terminals are connected by a + connection and one of them is the bottom terminal of its host oscillator, the other terminal is labelled the bottom terminal of its own host oscillator. If two terminals are connected by a - connection and one of them is the bottom terminal of its host oscillator, the other terminal is labelled the top terminal of its own host oscillator. Through this process, we can identify the bottom terminals of all the oscillators. The `bottom' labelling is shown in Fig. \ref{twospin} for $J_{ij}=1$ and $J_{ij}=-1$.         

Let the potential at the `bottom' terminal of oscillator $i$ be $V_{bi}$. The current that flows out from the bottom terminal of the $i$-th oscillator into the resistor that connects it to the $j$-th oscillator is $i_{ij1}$. Similarly, the current that flows out from the top terminal of the $i$-th oscillator into the resistor that connects it to the $j$-th oscillator is $i_{ij0}$. In the $i$-th oscillator, the voltage difference between the top and the bottom terminals of the capacitor is denoted by $V_i$, the current passing through the inductor from the top to the bottom terminals is $I_{li}$, and the current passing through the capacitor from the top to the bottom terminals is $I_{ci}$. All of this notation is again indicated in Fig. \ref{twospin}.

The circuit equations are:
\begin{alignat}{2}
&C_s\dot{V}_{si}=I_{3i}-I_{li}-\sum_jI_{ij0},\ \ &&C_p\dot{V}_{pi}=I_{2i}-I_{3i},\label{mult1}\\
&V_{si}=L_s\dot{I}_{li},\ \ &&V_{pi}=L_p\lp\dot{I}_{pi}-\dot{I}_{2i}\rp\label{mult2}
\end{alignat}
\begin{equation}
    I_{3i}=C_0\lp \dot{V}_{pi}-\dot{V}_{si}\rp+2C_N\lp V_{pi}-V_{si}\rp\lp \dot{V}_{pi}-\dot{V}_{si}\rp\label{mult3}
\end{equation}
\begin{gather}
    \sum_jI_{ij0}+\sum_jI_{ij1}=0,\ \text{for all }i\in\{1,2,\dots,N\}\label{mult4}\\
    V_{b1}=0\label{mult5}
\end{gather}
\begin{widetext}
\begin{equation}
    \begin{split}
    I_{ijk}&=k\ls\lp\frac{V_{bi}-V_{bj}}{R}\rp\lp\frac{1+J_{ij}}{2}\rp+\lp\frac{V_{bi}-V_{bj}-V_{sj}}{R}\rp\lp\frac{1-J_{ij}}{2}\rp\rs\\
    &+\lp1-k\rp\ls\lp\frac{V_{bi}+V_{si}-V_{bj}-V_{sj}}{R}\rp\lp\frac{1+J_{ij}}{2}\rp+\lp\frac{V_{bi}+V_{si}-V_{bj}}{R}\rp\lp\frac{1-J_{ij}}{2}\rp\rs,\\
    &\text{for all }i\in\{1,2,\dots,N\},\ \text{for all }j\in\{1,2,\dots,N\},\ j\ne i,\ \text{for all }k\in\{0,1\}
    \end{split}\label{mult6}
\end{equation}
Eqs. \eqref{mult1}, \eqref{mult2}, \eqref{mult3}, \eqref{mult4}, and \eqref{mult6} are the current law, voltage law, and device characteristics at different places in the circuit. Eq. \eqref{mult5} fixes the voltage reference by setting the potential of the bottom terminal of the first oscillator to $0$.

Eq. \eqref{mult6} yields:
\begin{equation}
    \sum_jI_{ij0}=\sum_{j:j\ne i}\ls\frac{V_{bi}-V_{bj}+V_{si}}{R}-\frac{V_{sj}}{R}\lp\frac{1+J_{ij}}{2}\rp\rs\label{mediate1}
\end{equation}
Next, plugging Eq. \eqref{mult6} into Eq. \eqref{mult4}, we get:
\begin{equation}
    \sum_{j:j\ne i}{2\lp\frac{V_{bi}-V_{bj}}{R}\rp}=-\frac{\lp N-1\rp V_{si}}{R}+\sum_{j:j\ne i}\frac{V_{sj}}{R}\label{mediate2}
\end{equation}
Finally, substituting Eq. \eqref{mediate2} into Eq. \eqref{mediate1}, we get:
\begin{equation}
    2R\lp \sum_jI_{ij0}\rp=\lp N-1\rp V_{si}-\sum_{j:j\ne i}J_{ij}V_{sj}
\end{equation}
We solve Eqs. \eqref{mult1} to \eqref{mult3} the same way as before, retaining only the $\omega_0$ and $2\omega_0$ terms in the signal and pump equations respectively, to obtain:
\begin{align}
    -\frac{N-1}{2R}\dot{V}_{si}+\frac{1}{2R}\sum_{j:j\ne i}J_{ij}\dot{V}_{sj}=&\frac{V_{si}}{L_s}+\lp C_0+C_s\rp\ddot{V}_{si}+4C_N\dot{V}_{pi}\dot{V}_{si}+2C_NV_{pi}\ddot{V}_{si}+2C_NV_{si}\ddot{V}_{pi}\label{coupledsignal}\\
    \dot{I}_{pi}=&\frac{V_{pi}}{L_p}+\lp C_0+C_p\rp\ddot{V}_{pi}+2C_N\dot{V}_{si}^2+2C_NV_{si}\ddot{V}_{si}\label{coupledpump}
\end{align}

\subsection{Slowly varying amplitude approximation}
Making the substitution $C_{0s}:=C_0+C_s$ and $C_{0p}:=C_0+C_p$ and plugging into Eq. \eqref{coupledsignal} the slowly varying amplitudes from above, we get the following signal equations:
\begin{equation}
\begin{split}
    \dot{A}_{si}-\frac{C_N\omega_0A_{pi}}{2C_{0s}}A_{si}+\frac{C_N}{2\omega_0C_{0s}}\ls2\dot{A}_{si}\dot{A}_{pi}-2\omega_0\dot{A}_{pi}B_{si}-2\omega_0\dot{B}_{si}A_{pi}\rs=&\frac{1}{2C_{0s}}\ls-\frac{N-1}{2R}\lp A_{si}-\frac{\dot{B}_{si}}{\omega_0}\rp\right.\\
    &\left.+\frac{1}{2R}\sum_{j:j\ne i}J_{ij}\lp A_{sj}-\frac{\dot{B}_{sj}}{\omega_0}\rp\rs
\end{split}\label{slowlycoupledcos}
\end{equation}
\begin{equation}
\begin{split}
    \dot{B}_{si}+\frac{C_N\omega_0A_{pi}}{2C_{0s}}B_{si}-\frac{C_N}{2\omega_0C_{0s}}\ls2\dot{B}_{si}\dot{A}_{pi}+2\omega_0\dot{A}_{pi}A_{si}+2\omega_0\dot{A}_{si}A_{pi}\rs=&\frac{1}{2C_{0s}}\ls-\frac{N-1}{2R}\lp B_{si}+\frac{\dot{A}_{si}}{\omega_0}\rp\right.\\
    &\left.+\frac{1}{2R}\sum_{j:j\ne i}J_{ij}\lp B_{sj}+\frac{\dot{A}_{sj}}{\omega_0}\rp\rs
\end{split}\label{slowlycoupledsin}
\end{equation}

As before, we drop Eq. \eqref{slowlycoupledsin} and all terms that contain $B_{si}$ in Eq. \eqref{slowlycoupledcos}, and apply the slowly-varying amplitude approximation to the remaining terms in Eq. \eqref{slowlycoupledcos} to obtain:
\begin{equation}
\begin{split}
    \dot{A}_{si}=\ls-\frac{N-1}{4RC_{0s}} A_{si}+\frac{1}{4RC_{0s}}\sum_{j:j\ne i}J_{ij} A_{sj}\rs+\frac{C_N\omega_0A_{pi}}{2C_{0s}}A_{si}
\end{split}\label{coupledcosfin}
\end{equation}
Eq. \eqref{coupledpump} reduces to:
\begin{equation}
    \dot{A}_{pi}=\frac{I_{pci}}{2C_{0p}}-\frac{C_N\omega_0A_{si}^2}{2C_{0p}}\label{coupledpumpfin}
\end{equation}

\subsection{Extension to cases where we do not have $J_{ij}=\pm1$ or all-to-all connections}\label{appsec2b}

So far, we have only considered $J$ matrices in which all the entries were chosen from $\{-1,1\}$. In this section, we describe the modifications required to generalize the coupled LC oscillator circuit when the $J$ values take on arbitrary real values expressed in binary form $\dots b_2b_1b_0.b_{-1}b_{-2}\dots$. We will let the sign of $J_{ij}$, positive or negative, be represented by $s_{ij}$. That is, $J_{ij}=s_{ij}\left|J_{ij}\right|$. The circuit equations Eqs. \eqref{mult1} to \eqref{mult5} carry over while Eq. \eqref{mult6} gets modified to:
\begin{equation}
    \begin{split}
    I_{ijk}&=k\ls\lp\frac{V_{bi}-V_{bj}}{R_{ij}}\rp\lp\frac{1+s_{ij}}{2}\rp+\lp\frac{V_{bi}-V_{bj}-V_{sj}}{R_{ij}}\rp\lp\frac{1-s_{ij}}{2}\rp\rs\\
    &+\lp1-k\rp\ls\lp\frac{V_{bi}+V_{si}-V_{bj}-V_{sj}}{R_{ij}}\rp\lp\frac{1+s_{ij}}{2}\rp+\lp\frac{V_{bi}+V_{si}-V_{bj}}{R_{ij}}\rp\lp\frac{1-s_{ij}}{2}\rp\rs,\\
    &\text{for all }i\in\{1,2,\dots,N\},\ \text{for all }j\in\{1,2,\dots,N\},\ j\ne i,\ \text{for all }k\in\{0,1\}
    \end{split}
\end{equation}
Following the same procedure as before, Eq. \eqref{mediate2} gets changed to:
\begin{equation}
    \sum_{j:j\ne i}{2\lp\frac{V_{bi}-V_{bj}}{R_{ij}}\rp}=-\sum_{j:j\ne i}\frac{V_{si}}{R_{ij}}+\sum_{j:j\ne i}\frac{V_{sj}}{R_{ij}}\label{mediate4}
\end{equation}
To implement an arbitrary $J_{ij}$, we use a `common' coupling resistor $R$, and binary multiples of it, $R_{m}=2^mR$. That is, $R_{-1}$ is $R/2$ and $R_{2}$ is $4R$. If $\left|J_{ij}\right|$ is written in binary form upto 3-bit precision as
\begin{equation}
    \left|J_{ij}\right|=b_12^1+b_02^0+b_{-1}2^{-1},
\end{equation}
$J_{ij}$ is implemented by setting:
\begin{equation}
    \frac{1}{R_{ij}}=\frac{b_1}{R_{-1}}+\frac{b_0}{R_{0}}+\frac{b_{-1}}{R_{1}}=\frac{1}{R}\left|J_{ij}\right|
\end{equation}
To see that this setting indeed does the job, we plug this expression for $R_{ij}$ into Eq. \eqref{mediate4} and complete the calculation to see that Eq. \eqref{coupledcosfin} gets changed to:
\begin{equation}
    \dot{A}_{si}=\ls-\frac{\sum_{j:j\ne i}\left|J_{ij}\right|}{4RC_{0s}} A_{si}+\frac{1}{4RC_{0s}}\sum_{j:j\ne i}J_{ij} A_{sj}\rs+\frac{C_N\omega_0A_{pi}}{2C_{0s}}A_{si}\label{coupledsignalnonbinJ}
\end{equation}
The pump equation Eq. \eqref{coupledpumpfin} remains unchanged.

\subsection{Including thermal noise in the coupling resistors}\label{appsec2c}
Thermal noise is incorporated into the circuit by adding noise voltage sources $V^{(n)}_{ij0}$ and $V^{(n)}_{ij1}$ in series with the coupling resistors $R_{ij0}$ and $R_{ij1}$ respectively. Further, the thermal noise generated by the internal resistors of the signal and pump LC oscillators is modeled by adding noise current sources $I^{(n)}_{si}$ and $I^{(n)}_{pi}$ in parallel to the two tanks respectively. Then, the counterparts of Eqs. \eqref{coupledsignal} and \eqref{coupledpump} are:
\begin{align}
\begin{split}
    -\frac{\sum_{j:j\ne i}|J_{ij}|}{2R}\dot{V}_{si}+\frac{1}{2R}\sum_{j:j\ne i}J_{ij}\dot{V}_{sj}-&\frac{1}{2R}\sum_{j:j\ne i}|J_{ij}|\dot{V}^{(n)}_{ij1}+\frac{1}{2R}\sum_{j:j\ne i}|J_{ij}|\dot{V}^{(n)}_{ij0}+\dot{I}^{(n)}_{si}\\
    =&\frac{V_{si}}{L_s}+\frac{\dot{V}_{si}}{R_s}+C_{0s}\ddot{V}_{si}+4C_N\dot{V}_{pi}\dot{V}_{si}+2C_NV_{pi}\ddot{V}_{si}+2C_NV_{si}\ddot{V}_{pi}
\end{split}\label{signalwithnoise}\\
    \dot{I}^{(n)}_{pi}+\dot{I}_{pi}=&\frac{V_{pi}}{L_p}+\frac{\dot{V}_{pi}}{R_p}+C_{0p}\ddot{V}_{pi}+2C_N\dot{V}_{si}^2+2C_NV_{si}\ddot{V}_{si}\label{pumpwithnoise}
\end{align}
\end{widetext}
The slowly-varying amplitude approximation restricts the above equations to small frequency windows around $\omega_0$ and $2\omega_0$ respectively which means that only band-pass filtered versions of the white noise terms $I^{(n)}_{pi}$, $I^{(n)}_{si}$, $V^{(n)}_{ij0}$, and $V^{(n)}_{ij1}$ are retained in the slowly-varying equations. If the impulse responses of band-pass filters centered about $\omega_0$ and $2\omega_0$ are $h_s(t)$ and $h_p(t)$ respectively, the convolution operator is represented by $\textasteriskcentered$, and the slowly varying cosine and sine noise amplitudes are represented by $A^{(n)}$ and $B^{(n)}$ with the appropriate subscripts (additional $I$ in superscript to represent currents), we have the following expressions for the filtered noise terms:
\begin{align*}
    h_s(t)\ast I^{(n)}_{si}(t)&=\frac{A^{(n, I)}_{si}-iB^{(n, I)}_{si}}{2}e^{i\lp\omega_0 t+\phi_s\rp}+\text{c.c}\\
    h_s(t)\ast v^{(n)}_{ij0}&=\frac{A^{(n)}_{ij0}-iB^{(n)}_{ij0}}{2}e^{i\lp\omega_0 t+\phi_s\rp}+\text{c.c}\\
    h_s(t)\ast v^{(n)}_{ij1}&=\frac{A^{(n)}_{ij1}-iB^{(n)}_{ij1}}{2}e^{i\lp\omega_0 t+\phi_s\rp}+\text{c.c}\\
    h_p(t)\ast I^{(n)}_{pi}&=\frac{A^{(n, I)}_{pi}-iB^{(n, I)}_{pi}}{2}e^{i\lp2\omega_0 t\rp}+\text{c.c}
\end{align*}
Using standard formulae and assuming that $A^{(n)}$ and $B^{(n)}$ are identically distributed but independent random processes, the 2-time correlation functions of the slowly-varying noise amplitudes are:
\begin{align}
    \langle A^{(n, I)}_{si}(0)A^{(n, I)}_{si}(\tau)\rangle&=\langle B^{(n, I)}_{si}(0)B^{(n, I)}_{si}(\tau)\rangle\label{autocorrfirst}\\
    &=\frac{4kT}{R_s}\frac{h_s(\tau)\ast h_s(-\tau)}{\cos{\lp\omega_0\tau\rp}}\\
    \langle A^{(n)}_{ij0}(0)A^{(n)}_{ij0}(\tau)\rangle&=\langle B^{(n)}_{ij0}(0)B^{(n)}_{ij0}(\tau)\rangle\\
    &=4kTR_{ij}\frac{h_s(\tau)\ast h_s(-\tau)}{\cos{\lp\omega_0\tau\rp}}\\
    \langle A^{(n)}_{ij1}(0)A^{(n)}_{ij1}(\tau)\rangle&=\langle B^{(n)}_{ij1}(0)B^{(n)}_{ij1}(\tau)\rangle\\
    &=4kTR_{ij}\frac{h_s(\tau)\ast h_s(-\tau)}{\cos{\lp\omega_0\tau\rp}}\\
    \langle A^{(n, I)}_{pi}(0)A^{(n, I)}_{pi}(\tau)\rangle&=\langle B^{(n, I)}_{pi}(0)B^{(n, I)}_{pi}(\tau)\rangle\\
    &=\frac{4kT}{R_p}\frac{h_p(\tau)\ast h_p(-\tau)}{\cos{\lp2\omega_0\tau\rp}}\label{autocorrlast}
\end{align}
The slowly-varying versions of Eqs. \eqref{signalwithnoise} and \eqref{pumpwithnoise} are then:
\begin{gather}
\begin{split}
    \dot{A}_{si}=-&\frac{\sum_{j:j\ne i}\left|J_{ij}\right|}{4RC_{0s}} A_{si}+\frac{1}{4RC_{0s}}\sum_{j:j\ne i}J_{ij} A_{sj}\\
    -&\frac{1}{2R_sC_{0s}}A_{si}+\frac{C_N\omega_0A_{pi}}{2C_{0s}}A_{si}\\
    +&\frac{A_{si}^{(n,I)}}{2C_{0s}}-\frac{1}{4RC_{0s}}\sum_{j:j\ne i}\lvert J_{ij}\rvert A^{(n)}_{ij1}\\
    +&\frac{1}{4RC_{0s}}\sum_{j:j\ne i}\lvert J_{ij}\rvert A^{(n)}_{ij0}
\end{split}\label{signalslowlynoise}\\
\dot{A}_{pi}=\frac{I_{pci}+A_{pi}^{(n,I)}}{2C_{0p}}-\frac{1}{2R_pC_{0p}}A_{pi}-\frac{C_N\omega_0A_{si}^2}{2C_{0p}}\label{pumpslowlynoise}
\end{gather}
If the signal and pump band-pass filters are assumed to be perfectly rectangular with unit real frequency response, the impulse responses $h_s(\tau)$ and $h_p(\tau)$ satisfy:
\begin{align}
    h_s(\tau)\ast h_s(-\tau)=h_s(\tau)&=\frac{2\sin{\lp\Delta\omega \tau/2\rp}}{\pi \tau}\cos{\lp\omega_0 \tau\rp}\\
    h_p(\tau)\ast h_p(-\tau)=h_p(\tau)&=\frac{2\sin{\lp\Delta\omega \tau/2\rp}}{\pi \tau}\cos{\lp2\omega_0 \tau\rp}
\end{align}
Next, we introduce noise processes $n_{si}(t)$ and $n_{pi}(t)$ to capture the noise terms in Eqs. \eqref{signalslowlynoise} and \eqref{pumpslowlynoise}: 
\begin{align}
    n_{si}&=2RA_{si}^{(n,I)}-\sum_{j:j\ne i}|J_{ij}|A_{ij1}^{(n)}+\sum_{j:j\ne i}|J_{ij}|A_{ij0}^{(n)}\label{noise1}\\
    n_{pi}&=A_{pi}^{(n,I)}\label{noise2}
\end{align}
Since the pump noise term Eq. \eqref{noise2} is straightforward to implement in the MATLAB \texttt{sde} solver, we shift our attention to the signal noise. In words, Eq. \eqref{noise1} tells us that the processes $n_{si}$, of which there are $N$, are linear combinations of $\mathcal{O}(N^2)$ independent Gaussian noise processes. We conclude from standard random process theory that the $n_{si}$ are Gaussian random processes too. This means they should be expressible as a linear combination of $N$ independent Gaussian noise processes $w_i$ instead of $\mathcal{O}(N^2)$ of them. This is a desirable representation because Eq. \eqref{signalslowlynoise} will then take the matrix-vector form:
\begin{equation}
    \dot{\bm{x}}=P\bm{x}+\frac{1}{4RC_{0s}}Q\bm{w}\label{sdemodel}
\end{equation}
for some matrices $P$ and $Q$. $P$ is readily extracted from Eq. \eqref{signalslowlynoise} whereas $Q$ is such that
\begin{equation}
    n_{si}=\sum_{j}Q_{ij}w_j\label{Qndef}
\end{equation}
Eq. \eqref{sdemodel} is a Langevin stochastic differential equation and can readily be simulated using MATLAB's \texttt{sde} function. 

We show next how to compute the matrix $Q$ that leads to correlations that are consistent with Eq. \eqref{noise1}. The 2-point correlation of $n_{si}(t)$ with itself (its autocorrelation) is:
\begin{equation}
\begin{split}
    \langle n_{si}(0)n_{si}(\tau)\rangle=&4R^2\langle A_{si}^{(n,I)}(0)A_{si}^{(n,I)}(\tau)\rangle\\
    &+2\sum_{j:j\ne i}|J_{ij}|^2\langle A_{ij0}^{(n)}(0)A_{ij0}^{(n)}(\tau)\rangle\\
    =8kTR&\left(\frac{2R}{R_s}+\sum_{j:j\ne i}|J_{ij}|\right)\frac{2\sin{(\Delta\omega\tau/2)}}{\pi\tau}
\end{split}\label{nsinsi}
\end{equation}
while the 2-point correlation between $n_{si}(t)$ and $n_{sk}(t)$ for $i\ne k$ is:
\begin{equation}
    \langle n_{si}(0)n_{sk}(\tau)\rangle=-8kTR|J_{ik}|J_{ik}\frac{2\sin{(\Delta\omega\tau/2)}}{\pi\tau}\label{nsinsk}
\end{equation}
Assuming that the processes $w_i$ have autocorrelation $2\sin{(\Delta\omega\tau/2)}/(\pi\tau)$ and using Eq. \eqref{Qndef}, we get:
\begin{align}
    \langle n_{si}(0)n_{si}(\tau)\rangle&=\lp\sum_jQ_{ij}^2\rp\frac{2\sin{(\Delta\omega\tau/2)}}{\pi\tau}\label{QiQi}\\
    \langle n_{si}(0)n_{sk}(\tau)\rangle&=\lp\sum_jQ_{ij}Q_{kj}\rp\frac{2\sin{(\Delta\omega\tau/2)}}{\pi\tau}\label{QiQk}
\end{align}
Equating the right hand sides of Eqs. \eqref{nsinsi} and \eqref{QiQi}, and those of Eqs. \eqref{nsinsk} and \eqref{QiQk}, we see that $Q$ is obtained by performing the Cholesky decomposition of a matrix $M$ constructed as follows:
\begin{align}
    M_{ik}&=-8kTR|J_{ik}|J_{ik},\ \text{if } i\ne k\\
    &=8kTR\left(\frac{2R}{R_s}+\sum_{j:j\ne i}|J_{ij}|\right),\ \text{if }i=k\label{Miequalk}
\end{align}
This completes the discussion of the signal noise.

At this point, we make a couple of comments on our code implementation. Firstly, our circuit had a cubic nonlinear saturating internal conductance $I=G_{\text{lin}}V+G_{\text{nonlin}}V^3$ in the signal circuit so the $2R/R_s$ term in Eq. \eqref{Miequalk} was replaced with $2R\lp G_{\text{lin}}+9G_{\text{nonlin}}A^2_{\text{sat}} \rp$. Secondly, we faced difficulties with generating band-limited white noise with sinc autocorrelation which is what the $w_i$ need to be. For this reason, we simply used pure white noise (Dirac delta autocorrelation) for the $w_i$. This assumption translates to forcing the slowly varying amplitudes to have Dirac delta autocorrelation instead of the sinc autocorrelation that was derived in Eqs. \eqref{autocorrfirst} to \eqref{autocorrlast}.   

\subsection{Extension to the case of non-zero local magnetic fields $h_i$}
In a more general form of the Ising problem, each spin also experiences a local magnetic field that adds to the total energy. The Hamiltonian is then:
\begin{equation}
    H=-\sum_ih_ix_i-\sum_{ij}J_{ij}x_ix_j
\end{equation}
This expression can be interpreted as an $N+1$ spin Ising Hamiltonian where $h_i/2$ are the $J_{i,N+1}$ connection coefficients of the first $N$ spins to a newly introduced $N+1$-th spin that is fixed to orientation $+1$. This viewpoint enables us to minimize this new Hamiltonian by simply adding an ac voltage source with phase corresponding to $+1$ to the original circuit and connecting it to the other oscillators through $h_i/2$ resistors in a manner exactly analogous to the $J$ resistors. The final signal equation of motion is:
\begin{equation}
\begin{split}
    \dot{A}_{si}=&\ls-\frac{|h_i|/2+
    \sum_{j}|J_{ij}|}{4RC_{0s}} A_{si}+\frac{1}{4RC_{0s}}\frac{h_{i}}{2} A_{\text{sat}}\right.\\
    &\left.+\frac{1}{4RC_{0s}}\sum_{j:j\ne i}J_{ij} A_{sj}\rs+\frac{C_N\omega_0A_{pi}}{2C_{0s}}A_{si}
\end{split}
\end{equation}

\section{Duality and the saddle point nature of $\lp\bm{x^*},\bm{\lambda^*}\rp$}
Let us say we are searching for \textbf{constrained global minima} instead of \textbf{constrained local minima}. The problem we are trying to solve is:
\begin{alignat*}{2}
    &\text{minimize}\ \ \ &&f(\bm{x})\\
    &\text{subject to }\ \ &&g_i(\bm{x})=0,\ i=1,\dots,p.
\end{alignat*}
Standard optimization textbooks show that this problem can be rewritten as:
\begin{equation}
    \min_{\bm{x}:\ g_i(x)=0\ \forall\ i}f(\bm{x})=\min_{\bm{x}}\lp\max_{\bm{\lambda}}L(\bm{x},\bm{\lambda})\rp
\end{equation}
where $L(\bm{x},\bm{\lambda})=f(\bm{x})+\sum_i\lambda_ig_i(\bm{x})$ is the Lagrange function. We have converted a constrained optimization problem into an \textbf{unconstrained} nested min-max optimization problem. The well-known min-max inequality that is true for arbitrary functions tells us that:
\begin{equation}
    \min_{\bm{x}}\lp\max_{\bm{\lambda}}L(\bm{x},\bm{\lambda})\rp\geq \max_{\bm{\lambda}}\lp\min_{\bm{x}}L(\bm{x},\bm{\lambda})\rp
\end{equation}
This relation holds for any optimization problem and is also called `weak duality'. For some special optimization problems\textemdash which includes many common convex optimization problems\textemdash we actually have equality:
\begin{equation}
    \min_{\bm{x}}\lp\max_{\bm{\lambda}}L(\bm{x},\bm{\lambda})\rp= \max_{\bm{\lambda}}\lp\min_{\bm{x}}L(\bm{x},\bm{\lambda})\rp\label{strongdual}
\end{equation}
The above relation says that the constrained global minimum $\bm{x^*}$ of $f(\bm{x})$ and its associated multiplier $\bm{\lambda^*}$ form a saddle point of $L(\bm{x},\bm{\lambda})$. To see why they form a saddle point of $L(\bm{x},\bm{\lambda})$, note that $\lp \bm{x},\argmax_{\bm{\lambda}}L(\bm{x},\bm{\lambda})\rp$ on the left-hand side represents a `1D' curved slice of the full space that passes through $\lp \bm{x^*},\bm{\lambda^*}\rp$. Moreover $L(\bm{x},\bm{\lambda})$ is minimized over this slice at $\lp \bm{x^*},\bm{\lambda^*}\rp$. Therefore, moving away from $\lp \bm{x^*},\bm{\lambda^*}\rp$ along the tangent to this slice increases $L$. Similarly, the right-hand side says that, over the `1D' curved slice represented by $\lp \argmin_{\bm{x}}L(\bm{x},\bm{\lambda}),\bm{\lambda}\rp$, $L(\bm{x},\bm{\lambda})$ is maximized at $\lp \bm{x^*},\bm{\lambda^*}\rp$. Therefore, moving away from $\lp \bm{x^*},\bm{\lambda^*}\rp$ along the tangent to this slice decreases $L$.

\section{Augmented Lagrange circuit using nonlinear resistors\textemdash Equations of Motion}\label{appsec4}
We insert a nonlinear resistor with the characteristic $I=G_0V+G_NV^3$ in parallel with all the signal circuit capacitors in the system to implement the cubic nonlinearity required by the Augmented Lagrange equations of motion. The circuit equations from before, Eqs. \eqref{mult1} to \eqref{mult6}, remain the same except for the first equation in Eq. \eqref{mult1} changing to:
\begin{equation}
    C_s\dot{V}_{si}=I_{3i}-I_{li}-\sum_jI_{ij0}-G_0V_{si}-G_NV_{si}^3
\end{equation}
Solving all the equations as before, the counterparts of Eqs. \eqref{coupledsignal} and \eqref{coupledpump} are:
\begin{align}
\begin{split}
    -\frac{N-1}{2R}\dot{V}_{si}+&\frac{1}{2R}\sum_{j:j\ne i}J_{ij}\dot{V}_{sj}=\frac{V_{si}}{L_s}+C_{0s}\ddot{V}_{si}+\\
    &4C_N\dot{V}_{pi}\dot{V}_{si}+2C_NV_{pi}\ddot{V}_{si}+2C_NV_{si}\ddot{V}_{pi}\\
    &+G_0\dot{V}_{si}+3G_NV_{si}^2\dot{V}_{si}
\end{split}\\
    \dot{I}_{pi}=&\frac{V_{pi}}{L_p}+C_{0p}\ddot{V}_{pi}+2C_N\dot{V}_{si}^2+2C_NV_{si}\ddot{V}_{si}\label{app4eq42}
\end{align}
Using the slowly-varying amplitude approximation and ignoring the sine components yields:
\begin{align}
\begin{split}
    \dot{A}_{si}=&\ls-\frac{N-1}{4RC_{0s}} A_{si}+\frac{1}{4RC_{0s}}\sum_{j:j\ne i}J_{ij} A_{sj}\rs+\frac{C_N\omega_0A_{pi}}{2C_{0s}}A_{si}\\
    &-\frac{G_0}{2C_{0s}}A_{si}-\frac{3G_N}{8C_{0s}}A_{si}^3
\end{split}\label{signalnonlin}\\
    \dot{A}_{pi}=&\frac{I_{pci}}{2C_{0p}}-\frac{C_N\omega_0A_{si}^2}{2C_{0p}}\label{pumpnonlin}
\end{align}

\section{Translating Quadratic Binary and MaxCut instances into Ising instances}\label{translate}
\subsection{Quadratic Binary to Ising}
The BiqMac collection specifies Quadratic Binary (0,1) \emph{minimization} problems by listing the coefficients $Q_{ij}$ of the terms $x_ix_j$ in the quadratic objective function. The coefficients form a symmetric matrix $Q$. The problem is stated precisely and recast as an Ising \emph{maximization} problem below:
\begin{align*}
    &\bm{x^*}=\argmin_{x_i\in\{0,1\}\forall i}\ \sum_{ij} Q_{ij}x_ix_j=\argmin_{x_i\in\{0,1\}\forall i}\ \bm{x^T}Q\bm{x}\\
    &=\argmax_{x_i=\pm1\forall i}\ \ls -\frac{1}{4}\bm{1^T}Q\bm{1}-\frac{1}{4}\sum_iQ_{ii}-\frac{1}{4}2\bm{1^T}Q\bm{x}-\frac{1}{4}\bm{x^T}\widetilde{Q}\bm{x}\rs\\
    &=\argmax_{x_i=\pm1\forall i}\ K+\frac{1}{4}\bm{h^Tx}+\frac{1}{4}\bm{x^T}J\bm{x}
\end{align*}
where $\widetilde{Q}$ is the same as matrix $Q$ but with the principal diagonal zeroed out, the effective Ising matrix $J_{ij}~:=~-~\widetilde{Q}_{ij}$, the effective Zeeman vector $\bm{h}~:=~-~2Q\bm{1}$, and the constant $K:=-\frac{1}{4}\bm{1^T}Q\bm{1}-\frac{1}{4}\sum_iQ_{ii}$.
\subsection{MaxCut to Ising}
The Gset collection specifies MaxCut problems by listing the edges $ij$ and their weights $w_{ij}$. The MaxCut optimization problem is stated and recast as an Ising problem below:
\begin{align*}
    \bm{x^*}&=\argmax_{x_i=\pm1\forall i}\ \frac{1}{8}\sum_{ij} w_{ij}\lp x_i-x_j\rp^2\\
    &=\argmax_{x_i=\pm1\forall i}\ \frac{1}{4}\sum_{ij} w_{ij}+\frac{1}{4}\sum_{ij} \lp-w_{ij}\rp x_ix_j\\
    &=\argmax_{x_i=\pm1\forall i}\ K+\frac{1}{4}\sum_{ij} J_{ij} x_ix_j
\end{align*}
where we introduced the effective Ising matrix $J_{ij}~:=~-~w_{ij}$
and the constant $K:=\frac{1}{4}\sum_{ij} w_{ij}$.

\section{Numerical results and parameter choices}\label{appsec6}
\subsection{Parameter choices}
\subsubsection{ODE and SDE solver settings}
All simulations were run for a total (circuit) time of $50\mu s$. The noiseless calculations were done using the \texttt{ode45} MATLAB solver while the noisy cases were run using the \texttt{sde} solver. The \texttt{ode45} solver adaptively picks time steps while a step size of $1ns$ was chosen for the \texttt{sde} calculations. 

The rms noise voltage across the signal capacitor in equilibrium is $V_{\text{noise}}=\sqrt{\frac{kT}{C_{0s}}}\approx5\mu V$ for the $C_{0s}$ chosen in the main text. All the signal circuit capacitor voltages at $t=0$ start out at this noise level in all our computations with the initial condition for the ode solver computations following a continuous uniform distribution between $-V_{\text{noise}}$ and $V_{\text{noise}}$ and the sde solver initial condition being chosen uniformly randomly from the discrete set $\lbr-V_{\text{noise}},V_{\text{noise}}\rbr$.

On the other hand, there was no randomness in our choice of the initial condition for the pump voltages. The initial pump voltage is the same for all the spins and is chosen such that the system experiences net gain right from $t=0$. From Eq. \eqref{coupledsignalnonbinJ}, the losses of the various oscillation modes of the circuit are proportional to the eigenvalues of the matrix $X$ whose elements are $X_{ij}=\delta_{ij}\lp\sum_{k:k\ne i}|J_{ik}|\rp-\lp1-\delta_{ij}\rp J_{ij}$ where $\delta_{ij}$ is the Kronecker delta. In the presence of a nonlinear saturating conductor, the losses increase further. We choose the initial pump voltage to create a gain that is a factor of 1.1 times larger than the 50th least loss in the system. If the $i$-th eigenvalue of a matrix $M$ is denoted by $\lambda_i(M)$, and $X$ is as defined earlier in this paragraph, our initial pump voltage for all oscillators in all computations is:
\begin{equation*}
    A_{pi}(0)=\frac{1}{C_N\omega_0}\times1.1\times\lp\frac{\lambda_{50}\lp X\rp}{2R}+G_0\rp
\end{equation*}
The nonlinear contribution to initial loss is ignored because all the signal amplitudes are initially at the noise level.

\subsubsection{Pump capacitance}
\begin{figure}[h]
\centering
\includegraphics[width=0.48\textwidth]{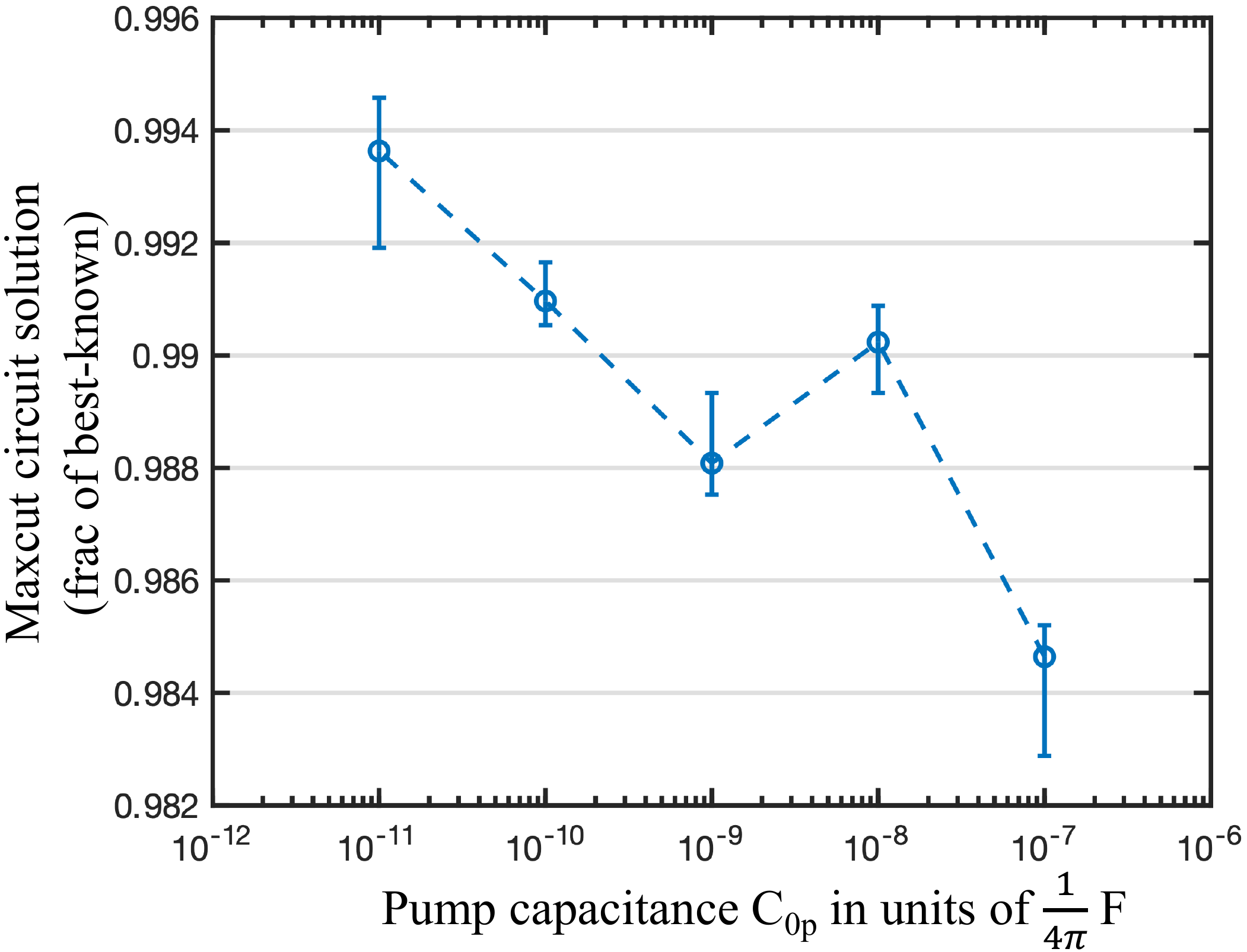}
\caption{Median, 25, and 75 percentile performance, as a fraction of the best-known solution, of 10 runs of the algorithm on the first Gset problem of size 800 for different values of pump capacitance.}
\label{fig:maxcutvscp}
\end{figure}

We recall from the Lagrange multipliers discussion in the main text that a good heuristic method to find constrained minima of optimization problems that satisfy only weak duality is to perform a fast gradient ascent in $\bm{\lambda}$ and a slow gradient descent in $\bm{x}$. Eq. \eqref{pumpnonlin} tells us that the speed of gradient ascent in the $A_{pi}$ (which are proportional to the Lagrange multipliers) directions is inversely proportional to $C_{0p}$. Therefore, reducing $C_{0p}$ should increase the speed of pump voltage evolution bringing the dynamics closer to the prescribed heuristic. This is demonstrated in Fig. \ref{fig:maxcutvscp} which shows that reducing the pump capacitance $C_{0p}$ does indeed improve the solution quality. To produce the results shown in the figure, the algorithm was run 10 times on the first Gset 800-spin problem for each value of $C_{0p}$ on the x-axis. The plot depicts the median, 25 and 75 percentiles of the 10 runs for each $C_{0p}$ as a fraction of the best-known solution for this problem. 

\begin{figure}[h]
\centering
\includegraphics[width=0.48\textwidth]{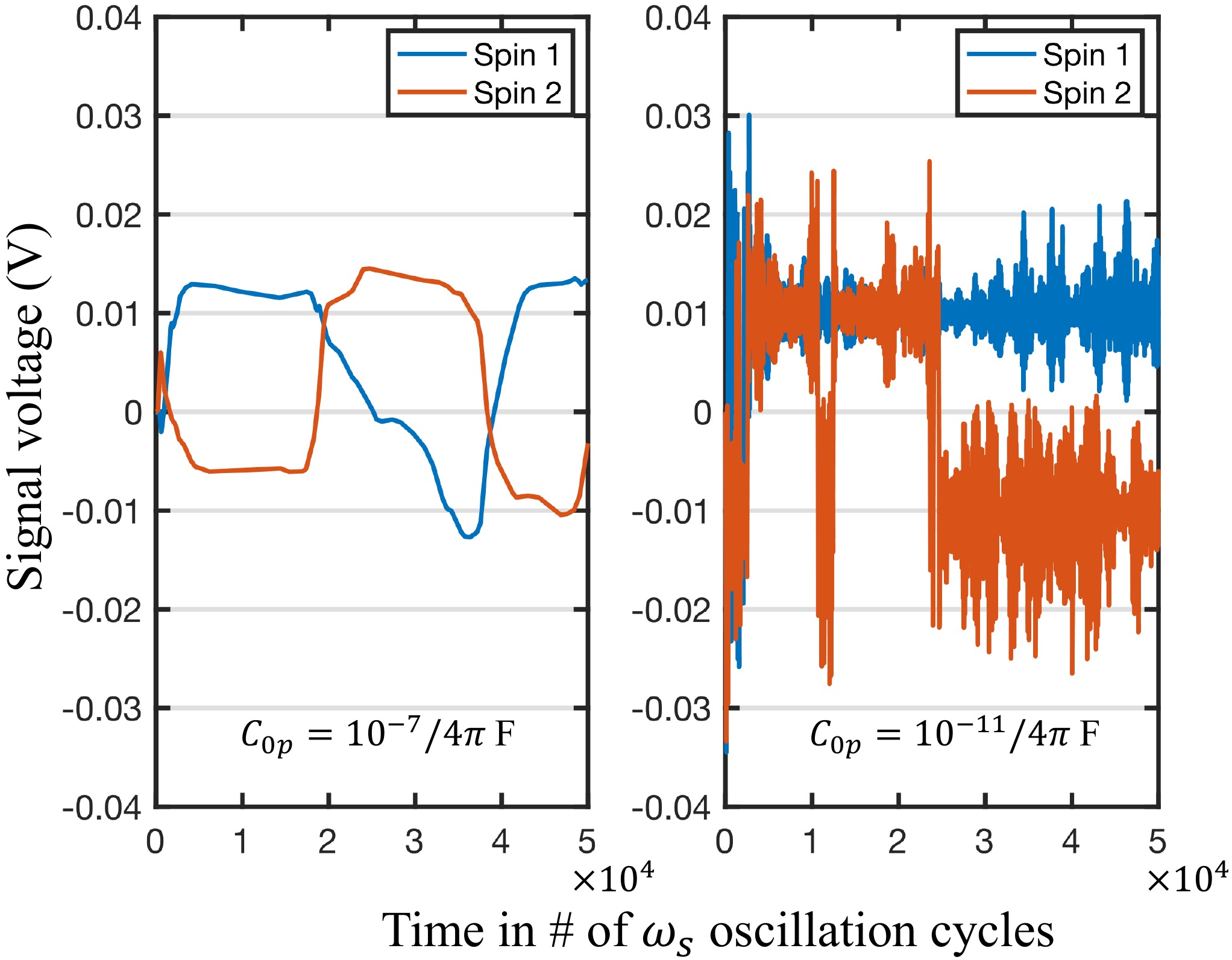}
\caption{Slow variation of signal voltage at high $C_{0p}$ shown in the left panel, fast variation at small $C_{0p}$ on the right.}
\label{fig:signalvstimecp}
\end{figure}

Reducing the pump capacitance $C_{0p}$ increases the speed with which the pump equation responds to deviations of the signal voltage from the saturation amplitude, and this in turn increases the speed of voltage variations in the signal circuit itself as shown in Fig. \ref{fig:signalvstimecp}. 

\begin{figure}[h]
\centering
\includegraphics[width=0.48\textwidth]{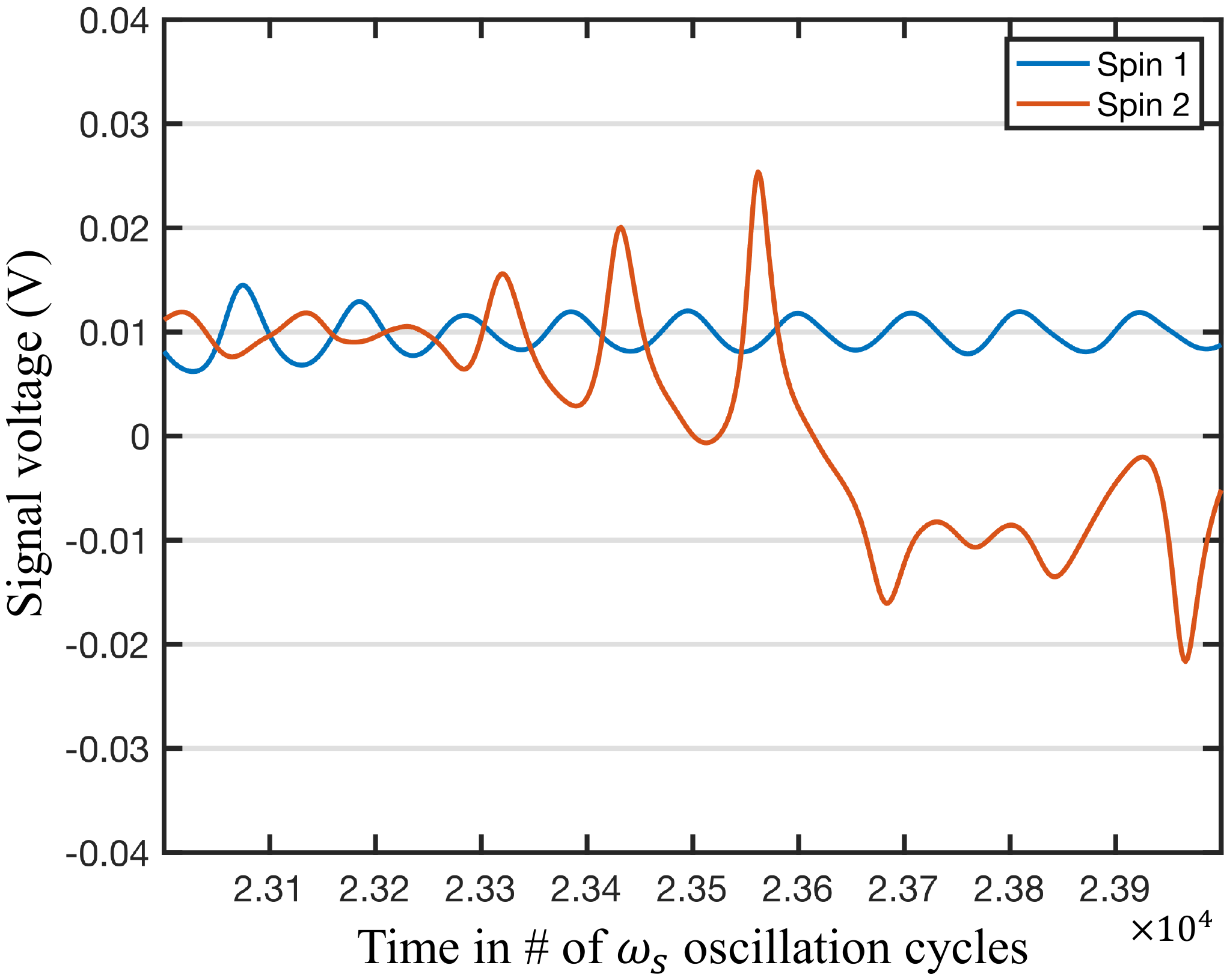}
\caption{Close up of the right panel of Fig. \ref{fig:signalvstimecp} showing that the slowly varying amplitude approximation is still valid.}
\label{fig:signalvstimecpzoom}
\end{figure}

We used $C_{0p}=10^{-11}/(4*\pi)$ in our simulations due to its better performance. One possible danger of using too small a $C_{0p}$ is that the fast variations it generates in the slowly varying amplitude could lead to a violation of the slowly varying amplitude approximation itself. Fig. ~\ref{fig:signalvstimecpzoom} zooms into the $C_{0p}=10^{-11}/(4*\pi)$ case and shows that the variation is on the range of hundreds of cycles, well within the validity regime of the slowly varying amplitude approximation.

\subsubsection{Effect of varying the strength of the internal nonlinear saturating conductor in the signal oscillators}
\begin{figure}[h]
\centering
\includegraphics[width=0.48\textwidth]{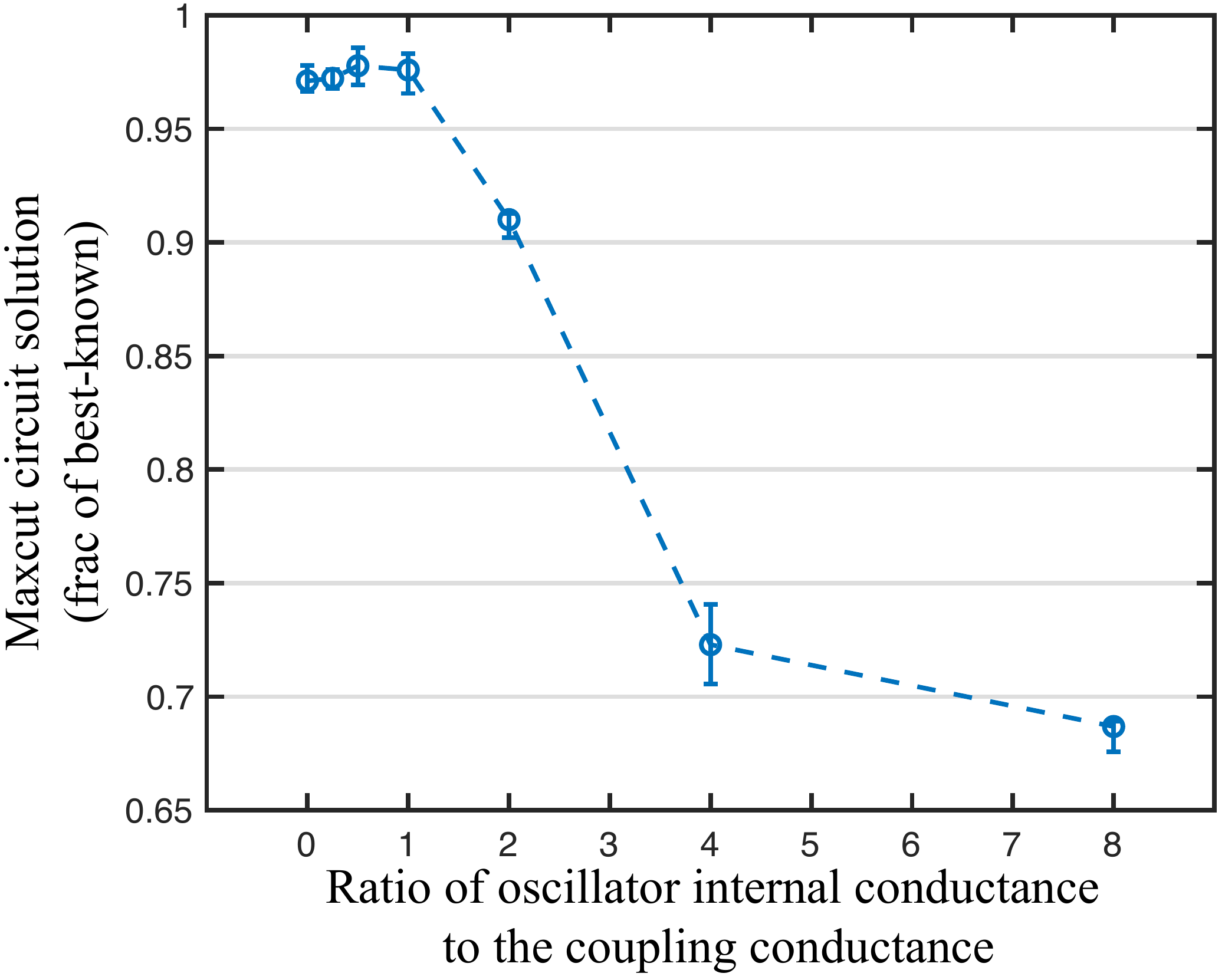}
\caption{Median, 25, and 75 percentile performance, as a fraction of the best-known solution, of 10 runs of the algorithm on the 6th Gset problem (size 800) for different values of signal oscillator internal conductance $G_0$.}
\label{fig:maxcutvsG0}
\end{figure}

\begin{figure}[h]
\centering
\includegraphics[width=0.48\textwidth]{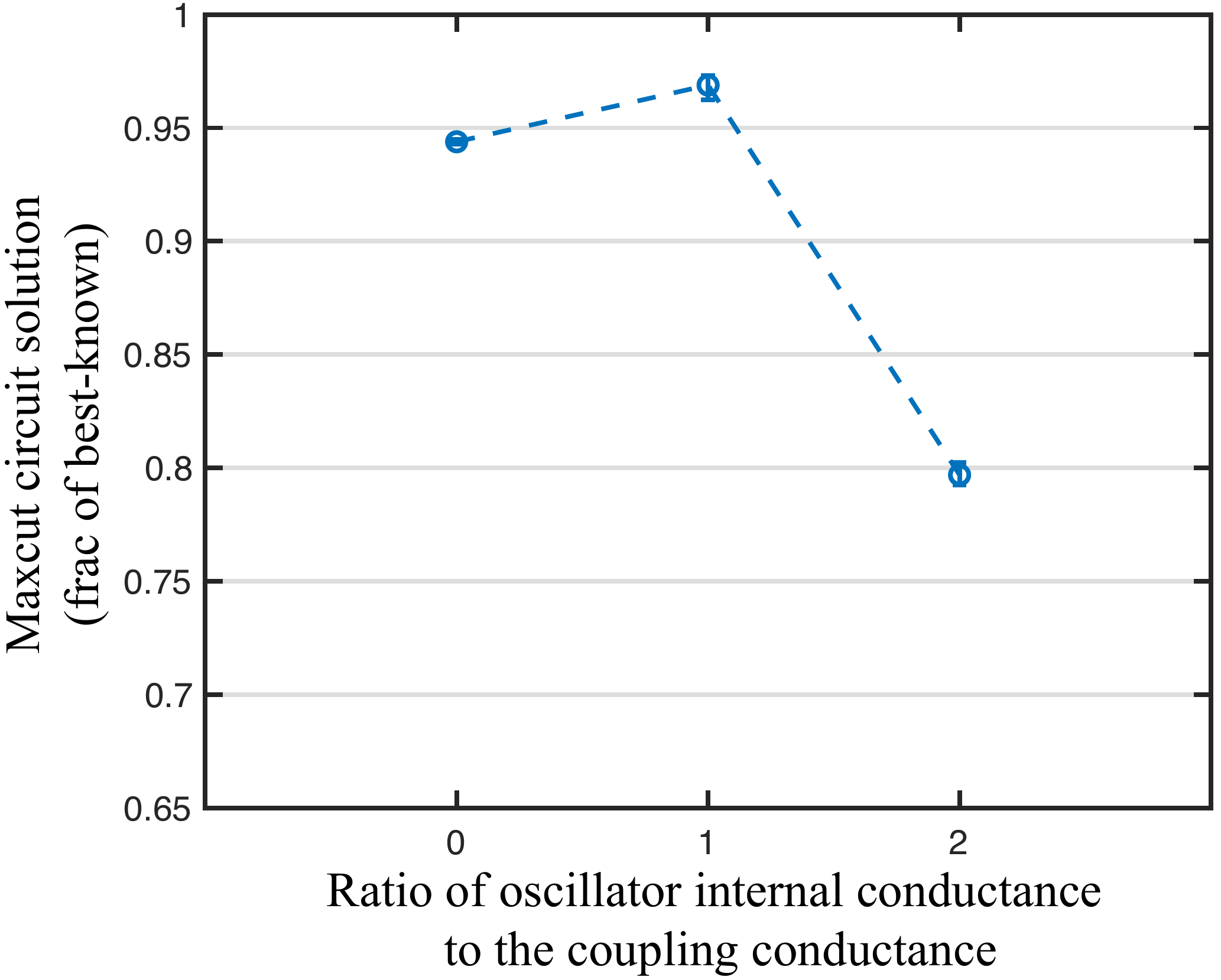}
\caption{Median, 25, and 75 percentile performance, as a fraction of the best-known solution, of 10 runs of the algorithm on the 27th Gset problem (size 2000) for different values of signal oscillator internal conductance $G_0$.}
\label{fig:maxcutvsG02000}
\end{figure}

The internal signal saturating conductor that implements the Augmented Lagrange method is:
\begin{equation}
    I=G_0V+G_NV^3
\end{equation}
$G_N$ is pegged to $G_0/A^2_{\text{sat}}$. This ensures that, once the signal amplitude reaches $A_{\text{sat}}$, the nonlinearity kicks in and limits the voltage. Scaling $G_0$ up increases the `steepness' of the nonlinear barrier faced by the signal voltage. Numerical simulations with $G_0=1/R$, where $R$ is the common coupling resistance, yielded performance that matched or bettered the no-nonlinearity performance for both 800 and 2000 spin Gset problems\textemdash this demonstrates that the Augmented Lagrange method is indeed better than the plain version. This is shown in Figs. \ref{fig:maxcutvsG0} and \ref{fig:maxcutvsG02000} and also Tables \ref{supptable1} and \ref{supptable2}. In Figs. \ref{fig:maxcutvsG0} and \ref{fig:maxcutvsG02000}, the x-axis shows the ratio $\frac{G_0}{1/R}$ while the plots themselves show the median, 25, and 75 percentile performance over 10 runs at each x-axis point as a fraction of the best-known solution. 

\subsubsection{Effect of varying the nonlinear capacitance $C_N$}
\begin{figure}[h]
\centering
\includegraphics[width=0.48\textwidth]{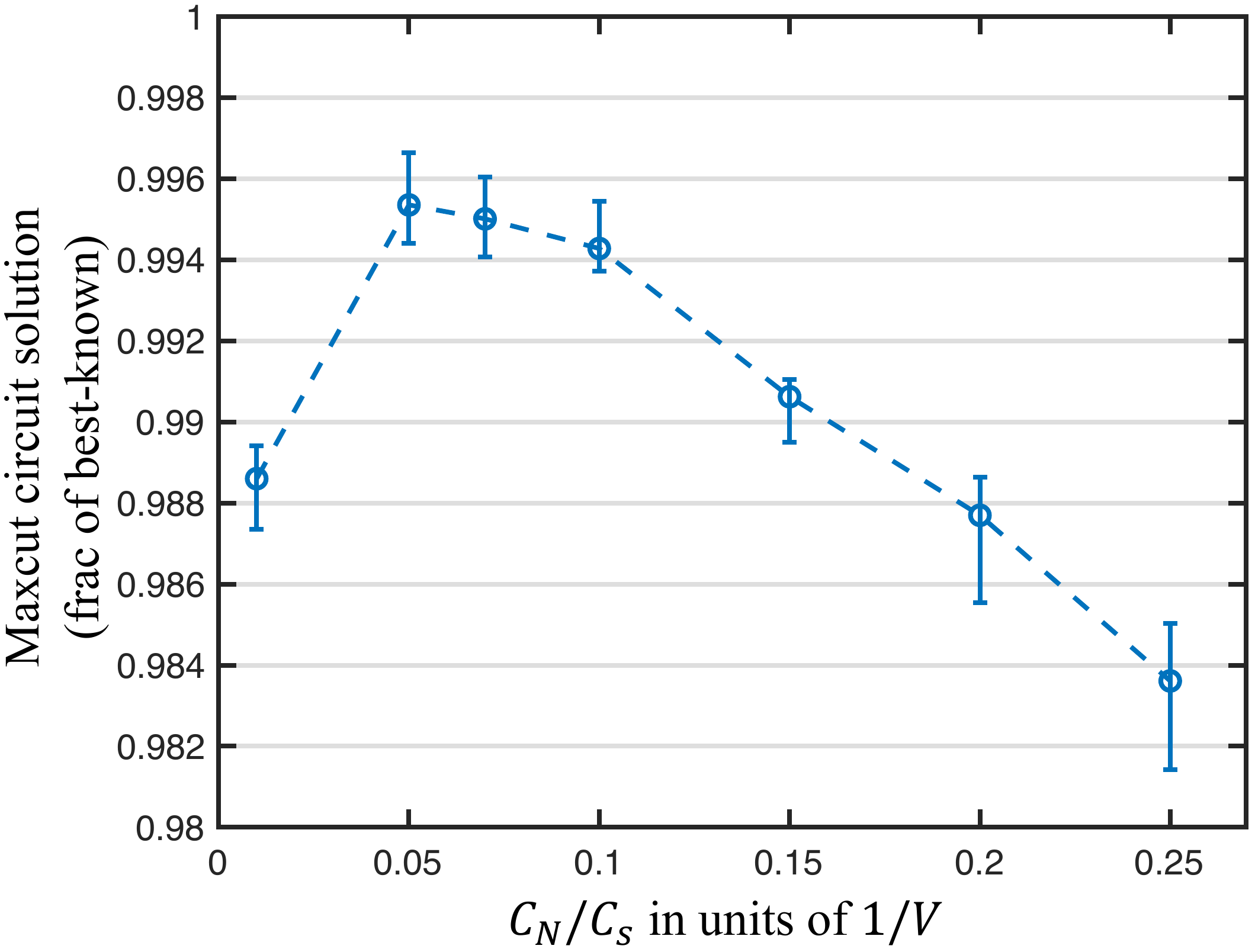}
\caption{Median, 25, and 75 percentile performance, as a fraction of the best-known solution, of 10 runs of the algorithm on the first Gset problem of size 800 for different values of nonlinear capacitance $C_N$. The y-axis ranges from 98\% to 100\% so the variation in performance is small.}
\label{fig:maxcutvsCN}
\end{figure}

The product of the nonlinear capacitance $C_N$ and the pump voltage $A_{pi}$ is the parametric gain of the $i$-th signal oscillator. Therefore, it is intuitive that varying $C_N$ should not have much effect because the pump voltage can compensate for the change. This is confirmed in Fig. \ref{fig:maxcutvsCN} where we see only small changes in the performance as $C_N$ is varied.  

\subsection{More results}\label{appsec6b}
Here we present results of the coupled oscillator Lagrange solver for Gset problems 1-10 (size 800), and problems 22-31 (size 2000). While Lagrange multipliers is outperformed by the Leleu approach, clever amalgamation of the two ideas could lead to better hybrid algorithms in the future.
\begin{table*}
\caption{Performance on Gset 1-10 (size 800) of the Goemans-Williamson (G-W) algorithm, the upper bound (UB G-W) implied by it, Leleu et al.'s approach, and the coupled oscillator approach without (Osc) and with (Osc NL) nonlinear resistors. The best and median are reported for 10 independent runs for the last two columns, while the results for Leleu are for 20 runs and are taken directly from their publication \cite{leleu_destabilization_2019}. Note that the ratio of G-W and UB G-W is not $\approx88\%$ in the last five rows because those problems have negative weight edges.}
\vspace{1mm}
\begin{tabular}{ ||c|c|c|c|c|c|c|| } 
\hline
\multirow{ 2}{*}{Prob} &\multirow{ 2}{*}{G-W} &\multirow{ 2}{*}{UB G-W} &\multirow{ 2}{*}{Metric} &\multirow{ 2}{*}{Leleu}  &\multirow{ 2}{*}{Osc} &\multirow{ 2}{*}{Osc NL}\\
& & & & & &\\
\hline\hline
 \multirow{ 2}{*}{1} &\multirow{ 2}{*}{11272} &\multirow{ 2}{*}{12838} &best &11624 & 11580 &11613\\
 & & &median &11624 &11552 &11558\\
 \hline
 \multirow{ 2}{*}{2} &\multirow{ 2}{*}{11277} &\multirow{ 2}{*}{12844} &best &11620 & 11575 &11596\\
  & & &median &11620 &11554 &11572\\
 \hline
 \multirow{ 2}{*}{3} &\multirow{ 2}{*}{11289} &\multirow{ 2}{*}{12857} &best &11622 &11588 &11586\\
  & & &median &11622 &11560 &11562\\
 \hline
 \multirow{ 2}{*}{4} &\multirow{ 2}{*}{11301} &\multirow{ 2}{*}{12871} &best &11646 &11611 &11641\\
  & & &median &11646 &11586 &11590\\
 \hline
 \multirow{ 2}{*}{5} &\multirow{ 2}{*}{11293} &\multirow{ 2}{*}{12862} &best &11631 &11591 &11578\\
  & & &median &11631 &11568 &11562\\
 \hline
 \multirow{ 2}{*}{6} &\multirow{ 2}{*}{1813} &\multirow{ 2}{*}{3387} &best &2178  & 2143 &2173\\
  & & &median &2178 &2124 &2144\\
 \hline
 \multirow{ 2}{*}{7} &\multirow{ 2}{*}{1652} &\multirow{ 2}{*}{3224} &best &2006 & 1975 &1973\\
  & & &median &2006 &1950 &1955\\
\hline
\multirow{ 2}{*}{8} &\multirow{ 2}{*}{1667} &\multirow{ 2}{*}{3243} &best &2005 &1966 &1992\\
  & & &median &2005 &1948 &1961\\
\hline
\multirow{ 2}{*}{9} &\multirow{ 2}{*}{1704} &\multirow{ 2}{*}{3278} &best &2054 &2010 &2043\\
  & & &median &2054 &1991 &2006\\
\hline
\multirow{ 2}{*}{10} &\multirow{ 2}{*}{1646} &\multirow{ 2}{*}{3218} &best &2000 &1956 &1979\\
  & & &median &2000 &1940 &1955\\
 \hline
\end{tabular}
\label{supptable1}
\end{table*}

\begin{table*}
\caption{Performance on the first 10 Gset 2000 vertex MaxCut problems of Leleu et al.'s approach, and the Osc and Osc NL approaches. Some of the problems have a `-' for the median in the Leleu column because the median could not be deduced from the published data.}
\vspace{1mm}
\begin{tabular}{ ||c|c|c|c|c|| } 
\hline
\multirow{ 2}{*}{Prob} &\multirow{ 2}{*}{Metric} &\multirow{ 2}{*}{Leleu}  &\multirow{ 2}{*}{Osc} &\multirow{ 2}{*}{Osc NL}\\
& & & &\\
\hline\hline
 \multirow{ 2}{*}{22} &best &13359 &13191 &13255\\
 &median &- &13176 &13231\\
 \hline
 \multirow{ 2}{*}{23} &best &13342 &13178 &13277\\
  &median &13342 &13151 &13228\\
 \hline
 \multirow{ 2}{*}{24} &best &13337 &13166 &13259\\
  &median &13337 &13150 &13232\\
 \hline
 \multirow{ 2}{*}{25} &best &13340 &13170 &13263\\
  &median &13340 &13154 &13228\\
 \hline
 \multirow{ 2}{*}{26} &best &13328 &13155 &13252\\
  &median &- &13142 &13228\\
 \hline
 \multirow{ 2}{*}{27} &best &3341  &3171 &3275\\
  &median &3341 &3156 &3237\\
 \hline
 \multirow{ 2}{*}{28} &best &3298 &3132 &3230\\
  &median &3298 &3112 &3185\\
\hline
\multirow{ 2}{*}{29} &best &3405 &3221 &3328\\
  &median &3405 &3206 &3302\\
\hline
\multirow{ 2}{*}{30} &best &3413 &3252 &3332\\
  &median &- &3226 &3287\\
\hline
\multirow{ 2}{*}{31} &best &3310 &3144 &3223\\
  &median &- &3125 &3203\\
 \hline
\end{tabular}
\label{supptable2}
\end{table*}

\end{document}